\documentclass[twocolumn, secnumarabic, amssymb, nobibnotes, prl, superscriptaddress]{revtex4}

\usepackage{graphicx} 
\usepackage{amsmath}
\usepackage{float}
\usepackage{color}
\usepackage[english]{babel}
\usepackage{array}
\usepackage[T1]{fontenc}
\usepackage[usenames,dvipsnames]{xcolor}
\usepackage{setspace}
\usepackage{hyperref}
\usepackage{bm}
\usepackage{lipsum} 
\usepackage{tikz}
\usetikzlibrary{math}
\usetikzlibrary{tikzmark}
\usepackage{stmaryrd}

\definecolor{color1}{rgb}{0.122,0.467,0.706}
\definecolor{color2}{rgb}{0.839,0.153,0.157}

\definecolor{color_RT1}{rgb}{1.0,0.455,0.0}
\definecolor{color_RT2}{rgb}{0.82,0.09,0.0}
\definecolor{color_RT3}{rgb}{0.545,0.0,0.0}
\definecolor{color_RT4}{rgb}{0.271,0.0,0.0}

\makeatletter
\renewcommand*{\fnum@figure}{{\normalfont \small{FIG.}~\thefigure}}
\makeatother

\begin{document}

\title{Tension-controlled switch between collective actuations in active solids}

\author{Paul Baconnier}
\affiliation{UMR CNRS Gulliver 7083, ESPCI Paris, PSL Research University, 75005 Paris, France.}
\author{Dor Shohat}
\affiliation{UMR CNRS Gulliver 7083, ESPCI Paris, PSL Research University, 75005 Paris, France.}
\affiliation{School of Physics and Astronomy, Tel-Aviv University, Tel Aviv 69978, Israel.}
\author{Olivier Dauchot}
\affiliation{UMR CNRS Gulliver 7083, ESPCI Paris, PSL Research University, 75005 Paris, France.}
%\date{\today}

\begin{abstract}
The recent finding of collective actuation in active solids, namely solids embedded with active units, opens the path towards multifunctional materials with genuine autonomy. In such systems, collective dynamics emerge spontaneously and little is known about the way to control or drive them. Here, we combine the experimental study of centimetric model active solids, the numerical study of an agent based model and theoretical arguments to reveal how mechanical tension can serve as a general mechanism for switching between different collective actuation regimes in active solids. We further show the existence of a hysteresis when varying back and forth mechanical tension, highlighting the non-trivial selectivity of collective actuations. 
\end{abstract}

\maketitle

One important aspect of metamaterial design is multi-functionality~\citep{bertoldi2017flexible, kim2018printing, siefert2019bio, liarte2020multifunctional, bossart2021oligomodal} --- the ability of a system to perform several different tasks. Multifunctional materials are usually actuated from an external source of work, which allows for a good control of the targeted functions~\citep{kim2018printing, siefert2019bio, bossart2021oligomodal}. Active matter, the individual components of which perform work, can also be tamed to actuate built-in functions autonomously~\citep{yuan2021recent}. It is therefore a promising alternative framework to create multifunctional materials with bona fide autonomy. Active solids ~\citep{koenderink2009active, henkes2011active, menzel2013traveling, berthier2013non, ferrante2013elasticity, prost2015active, ronceray2016fiber, woodhouse2018autonomous, briand2018spontaneously, giavazzi2018flocking, klongvessa2019active, maitra2019oriented, scheibner2020odd, baconnier2022selective} hold promise for such multifunctionality, especially when they exhibit emergent collective dynamics which selectively actuate only few deformation modes~\citep{henkes2011active, ferrante2013elasticity, woodhouse2018autonomous, baconnier2022selective}.

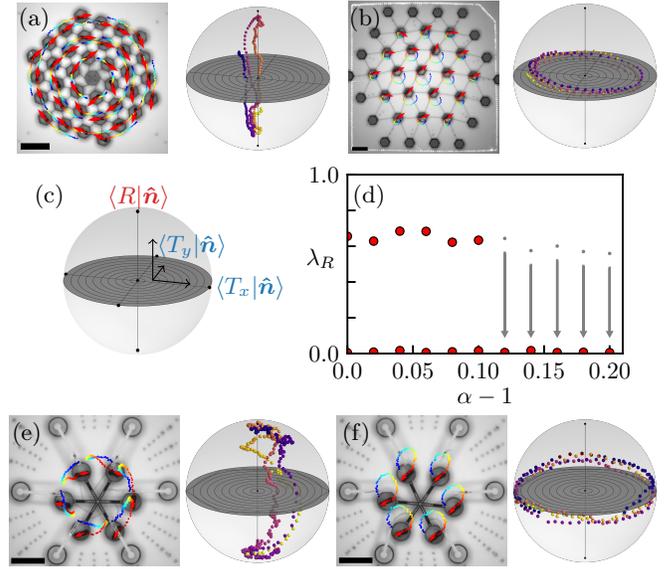
\begin{figure}[t!]
\centering
%\vspace*{-0.0cm}
%\hspace*{-0.45cm}
\begin{tikzpicture}

\tikzmath{ \x = 1.8; \y = 4.4; \dz = 1.2; \dzx=0.6; \dx=1.4; \xx = 3.125; \yy = 5.53; \xxx = 1.55; \yyy = 4.3; }
%\draw [->,thick] (\w/2, 0) -- (\w/2+\x,0);

\node[anchor=south west] at (1.09,1.6)
{\includegraphics[height=2.0cm]{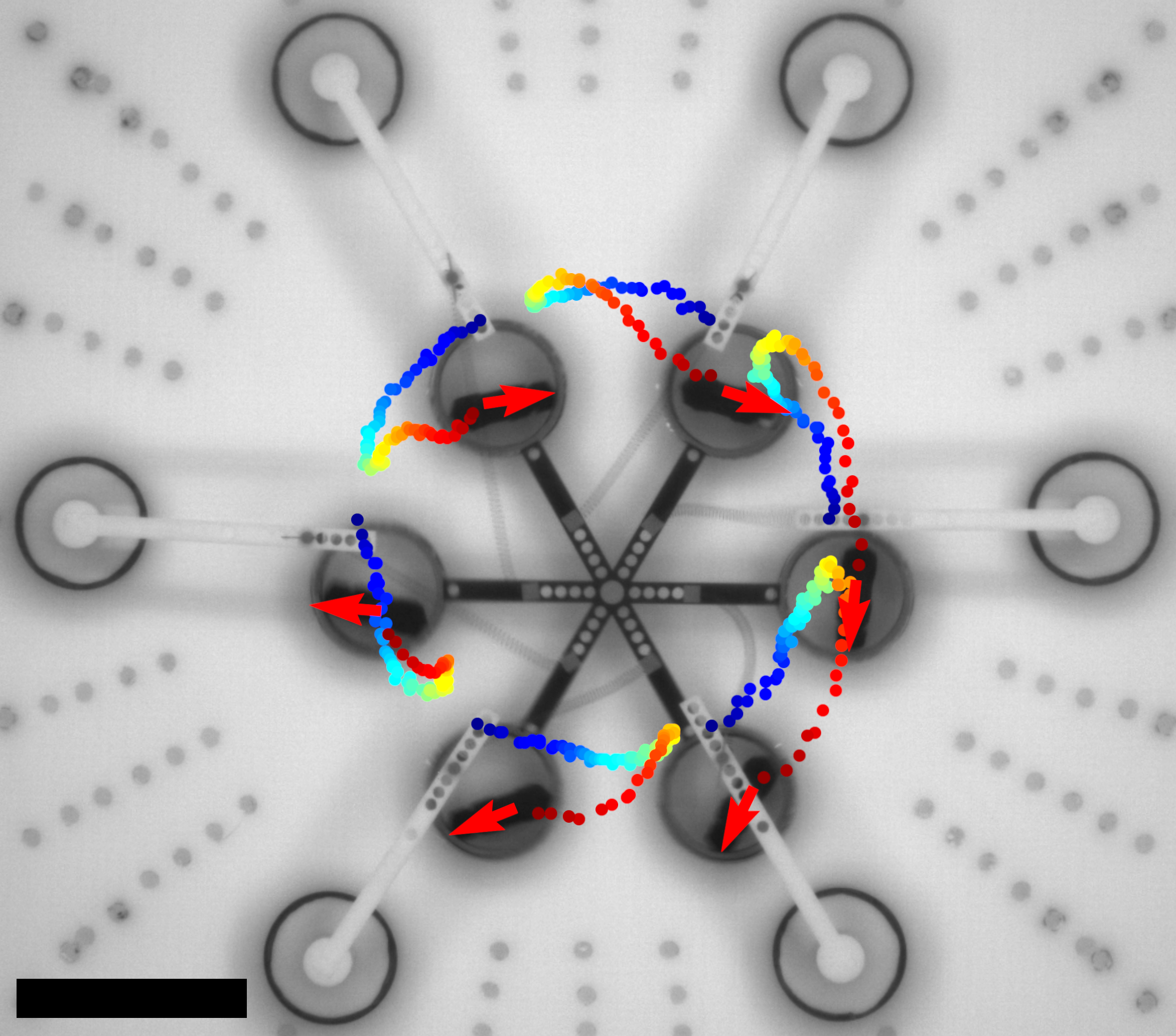}};
\node[anchor=south west] at (5.45,1.6)
{\includegraphics[height=2.0cm]{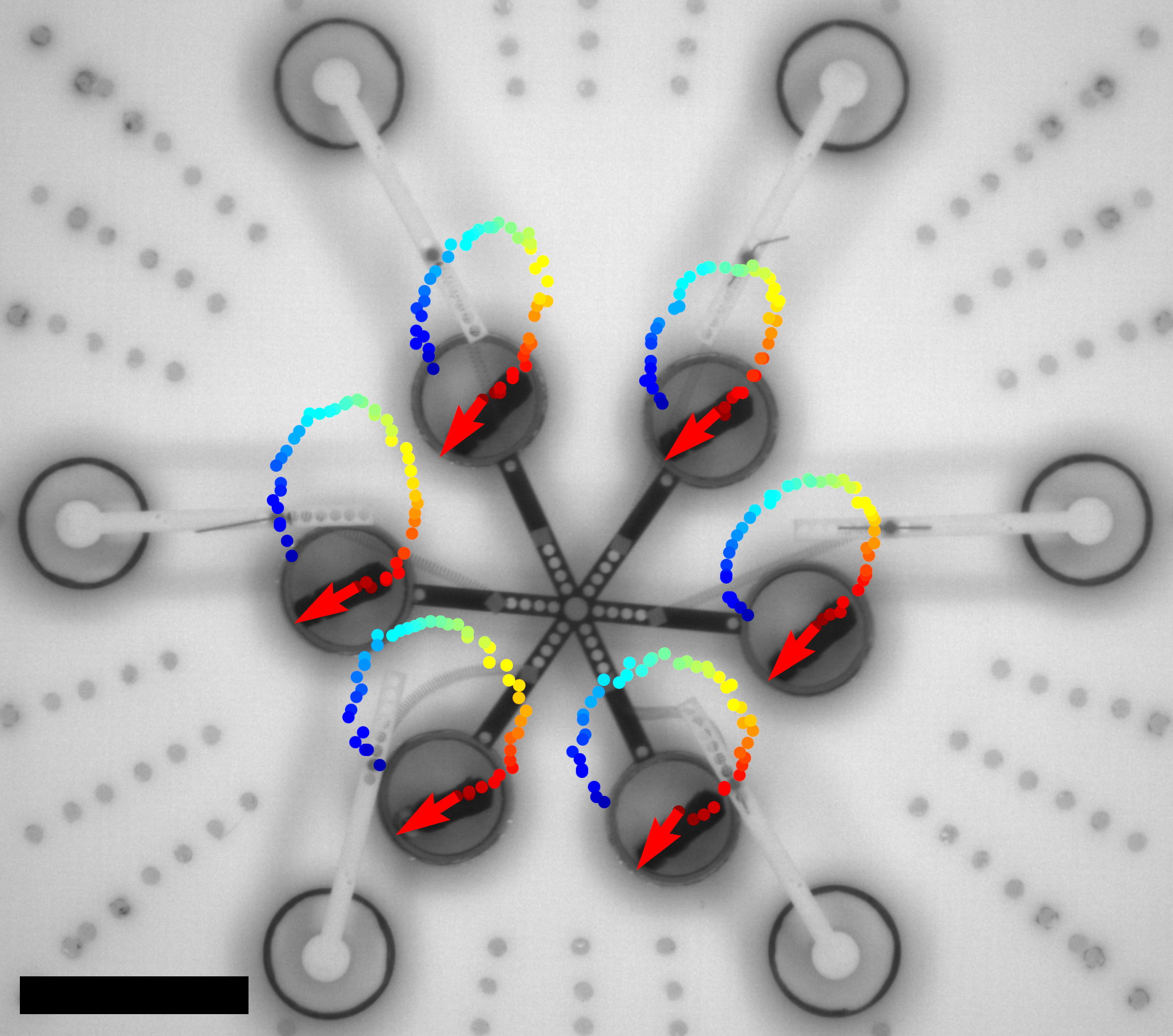}};

\node[anchor=south west] at (7.75,1.6)
{\includegraphics[height=2.0cm]{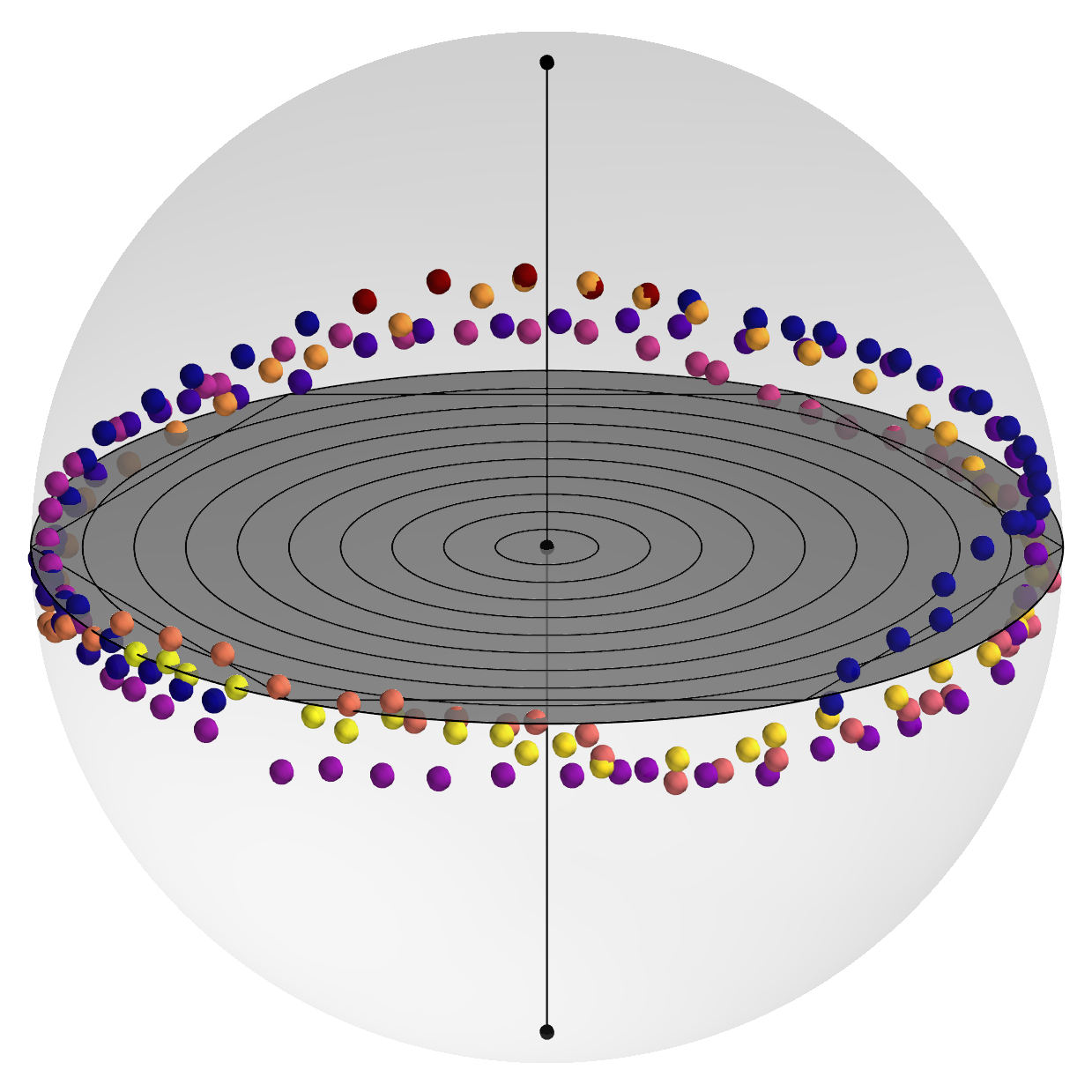}};
\node[anchor=south west] at (3.4,1.6)
{\includegraphics[height=2.0cm]{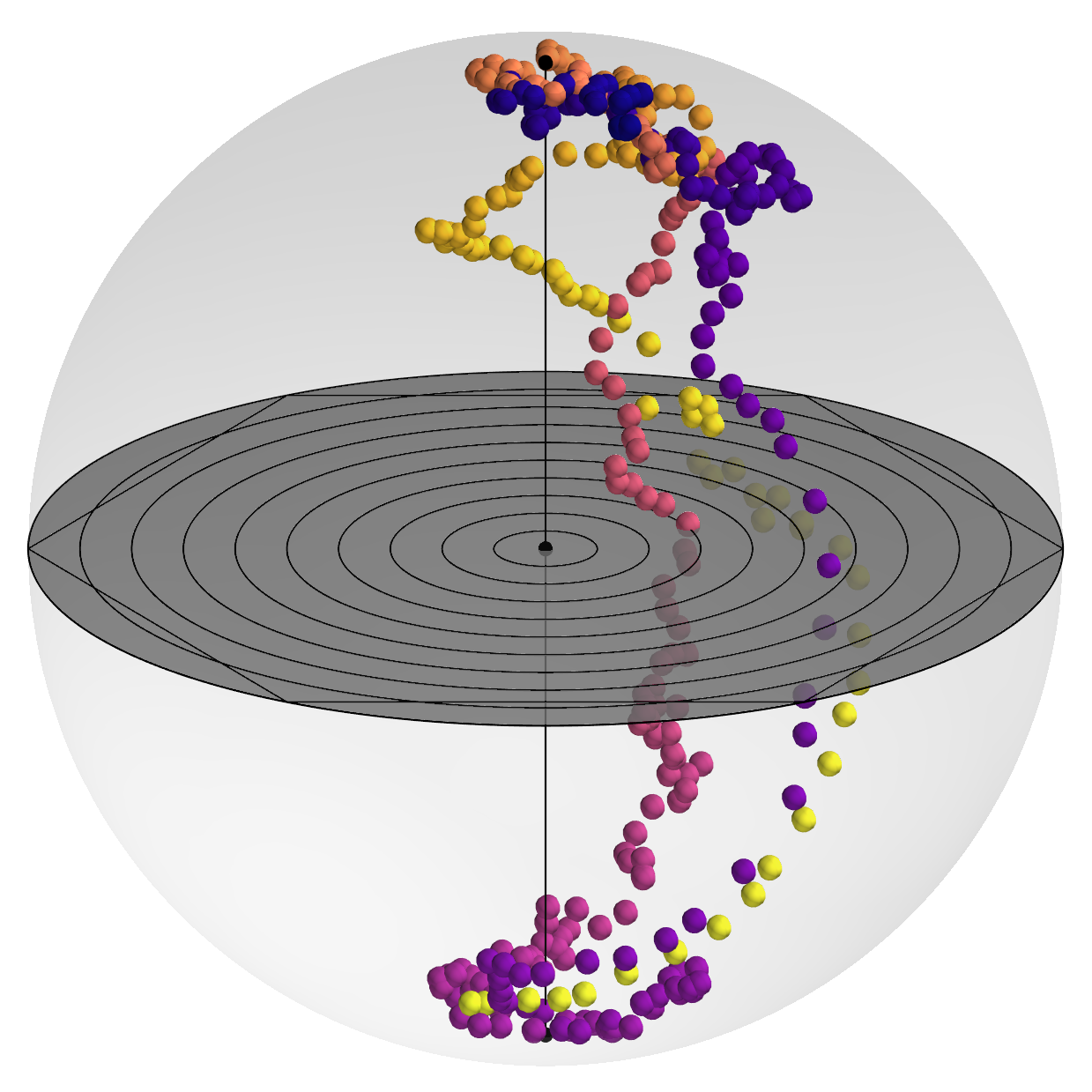}};

\node[anchor=south west] at (\x,\y)
{\includegraphics[width=2.0cm]{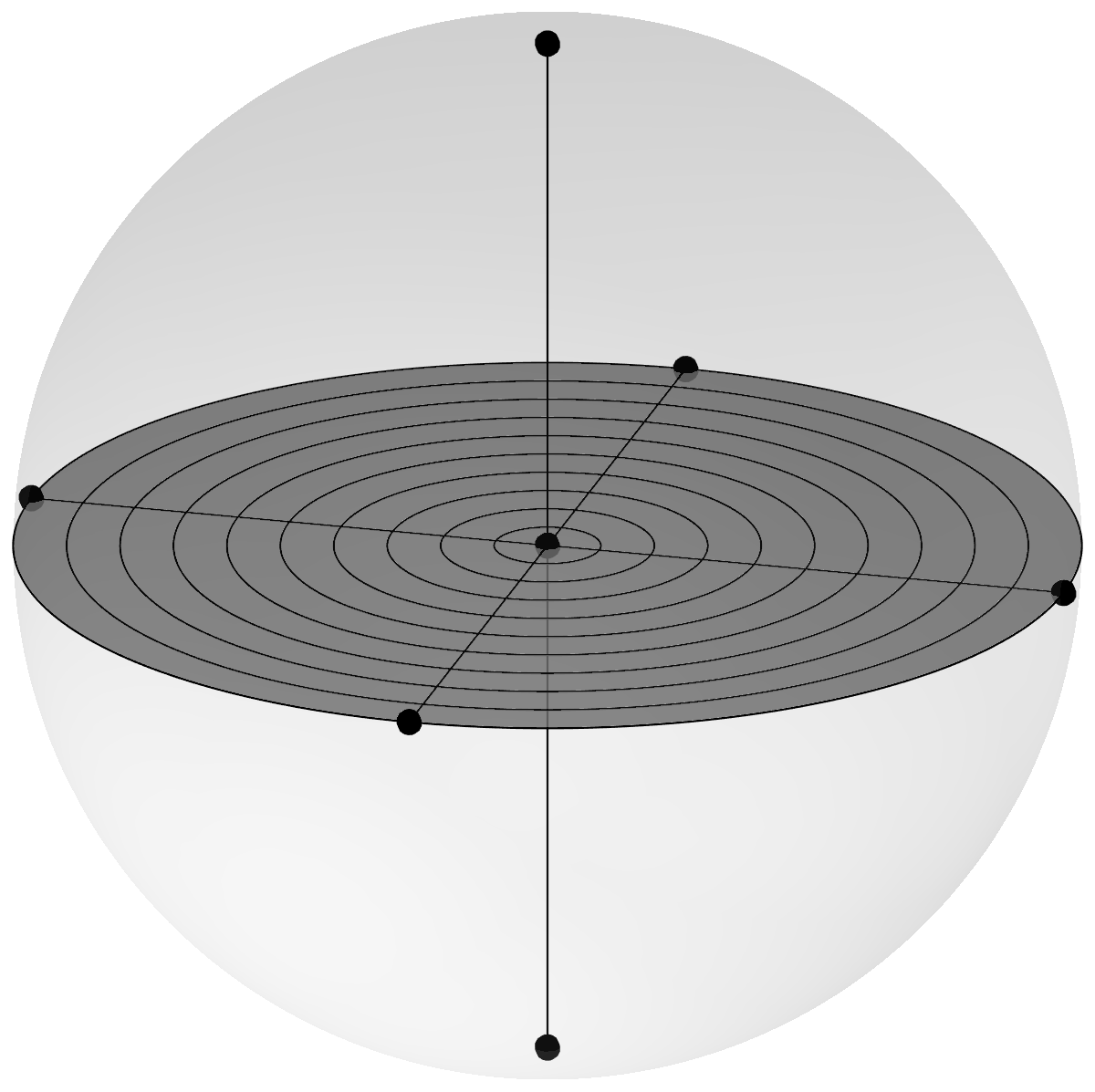}};

\node[anchor=south west] at (\x+\dzx,\y+0.75+\dz) {\small $\color{color2} \langle R | \boldsymbol{\hat{n}} \rangle$};
%\node[anchor=south west] at (\x+\dzx-0.2,\y+0.82-\dz) {\small $- \! \cowlor{color2} | R \rangle$};

\node[anchor=south west] at (\x+\dx+0.63,\y+0.75) {\small $\color{color1} \langle  T_x | \boldsymbol{\hat{n}} \rangle$};
%\node[anchor=south west] at (\x-\dx+0.63,\y+0.95) {\small $- \! \color{color1} | T_x \rangle$};

\node[anchor=south west] at (\x+0.63+0.63,\y+1.3) {\small $\color{color1} \langle  T_y | \boldsymbol{\hat{n}} \rangle$};
%\node[anchor=south west] at (\x+0.63-0.45,\y+0.25) {\small $- \! \color{color1} | T_y \rangle$};

%\filldraw[black] (\xx,\yy) circle (0.03);
\draw[->] (\xx,\yy) -- (\xx,\yy+0.55);
\draw[->] (\xx,\yy) -- (\xx+0.15,\yy+0.195);
\draw[->] (\xx,\yy) -- (\xx+0.5,\yy-0.05);

%\filldraw[black] (\xxx,\yyy) circle (0.03);
%\draw[->] (\xxx,\yyy) -- (\xxx,\yyy+0.55);
%\draw[->] (\xxx,\yyy) -- (\xxx+0.15,\yyy+0.195);
%\draw[->] (\xxx,\yyy) -- (\xxx+0.5,\yyy-0.05);

%\node[anchor=north east] at (\xxx+0.05,\yyy+0.05) {\footnotesize $\boldsymbol{O}$};
%\node[] at (\xxx+0.8,\yyy-0.1) {\small $a_{\color{color1} T_x}$};
%\node[] at (\xxx,\yyy+0.75) {\small $a_{\color{color2} R}$};
%\node[] at (\xxx+0.4,\yyy+0.3) {\small $a_{\color{color1} T_y}$};

%\node[] at (2.5,6.5) {\footnotesize $| \boldsymbol{\hat{n}} \rangle (t)$};

\node[anchor=south west] at (5.05,3.75)
{\includegraphics[height=3.4cm]{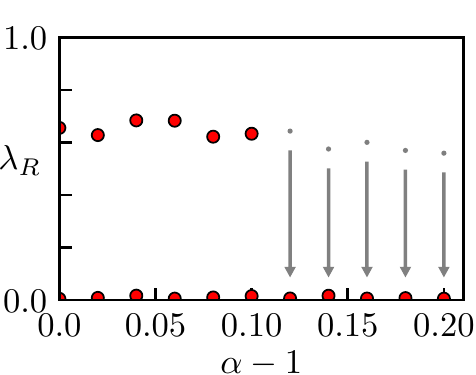}};

\node[anchor=south west] at (1.2,7.1)
{\includegraphics[width=2.0cm]{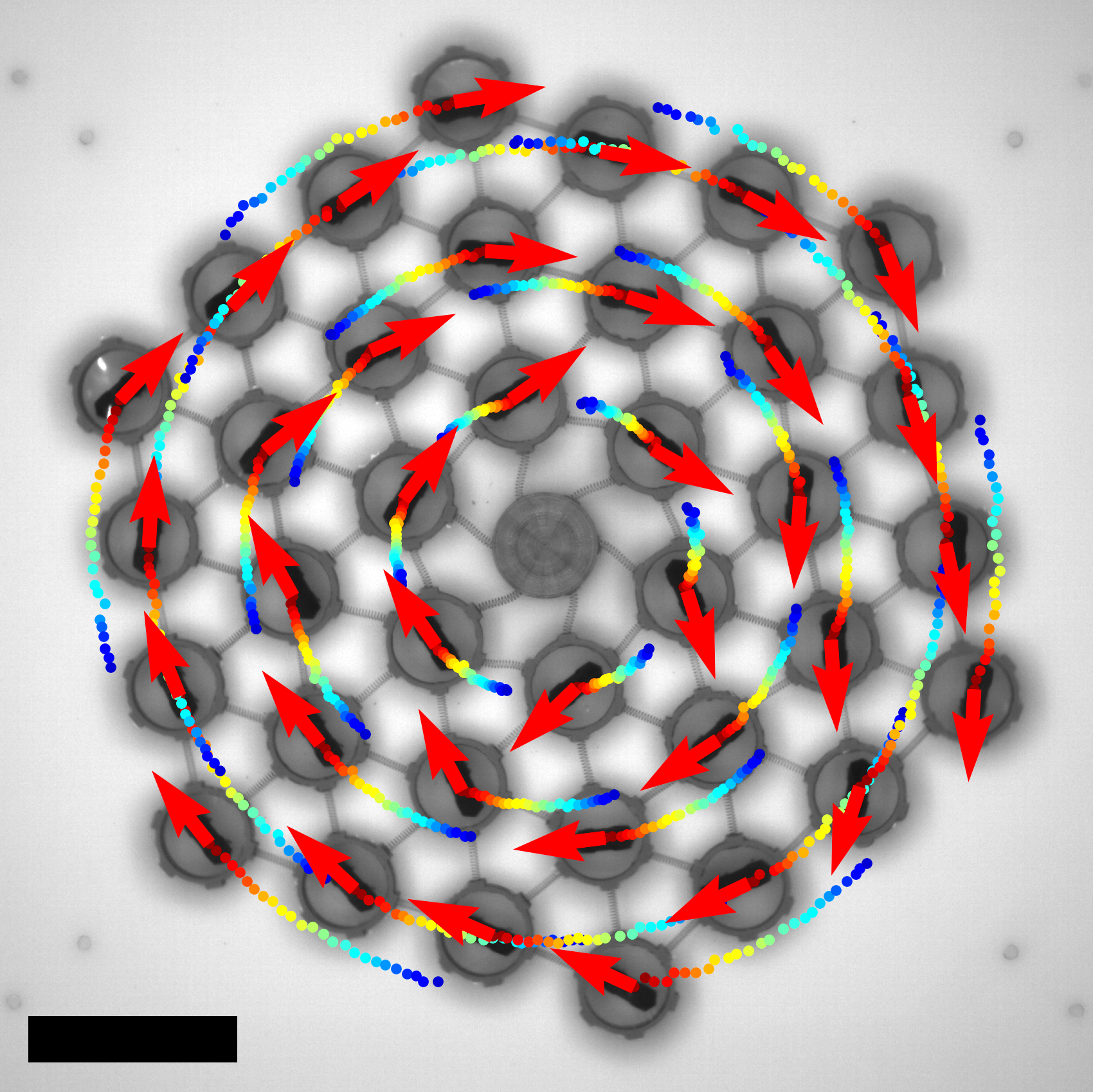}};
\node[anchor=south west] at (3.4,7.1)
{\includegraphics[width=2.0cm]{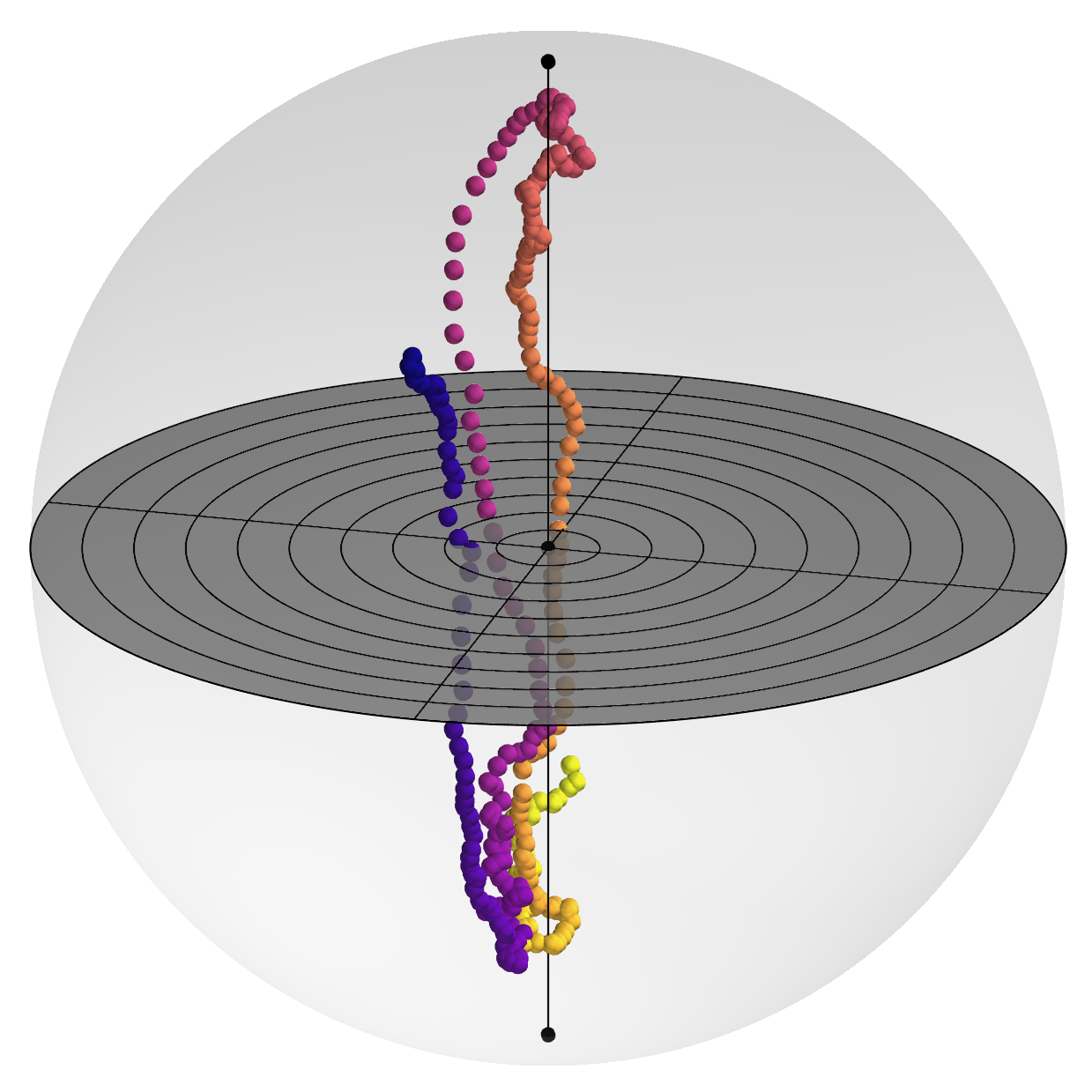}};
\node[anchor=south west] at (5.6,7.1)
{\includegraphics[width=2.0cm]{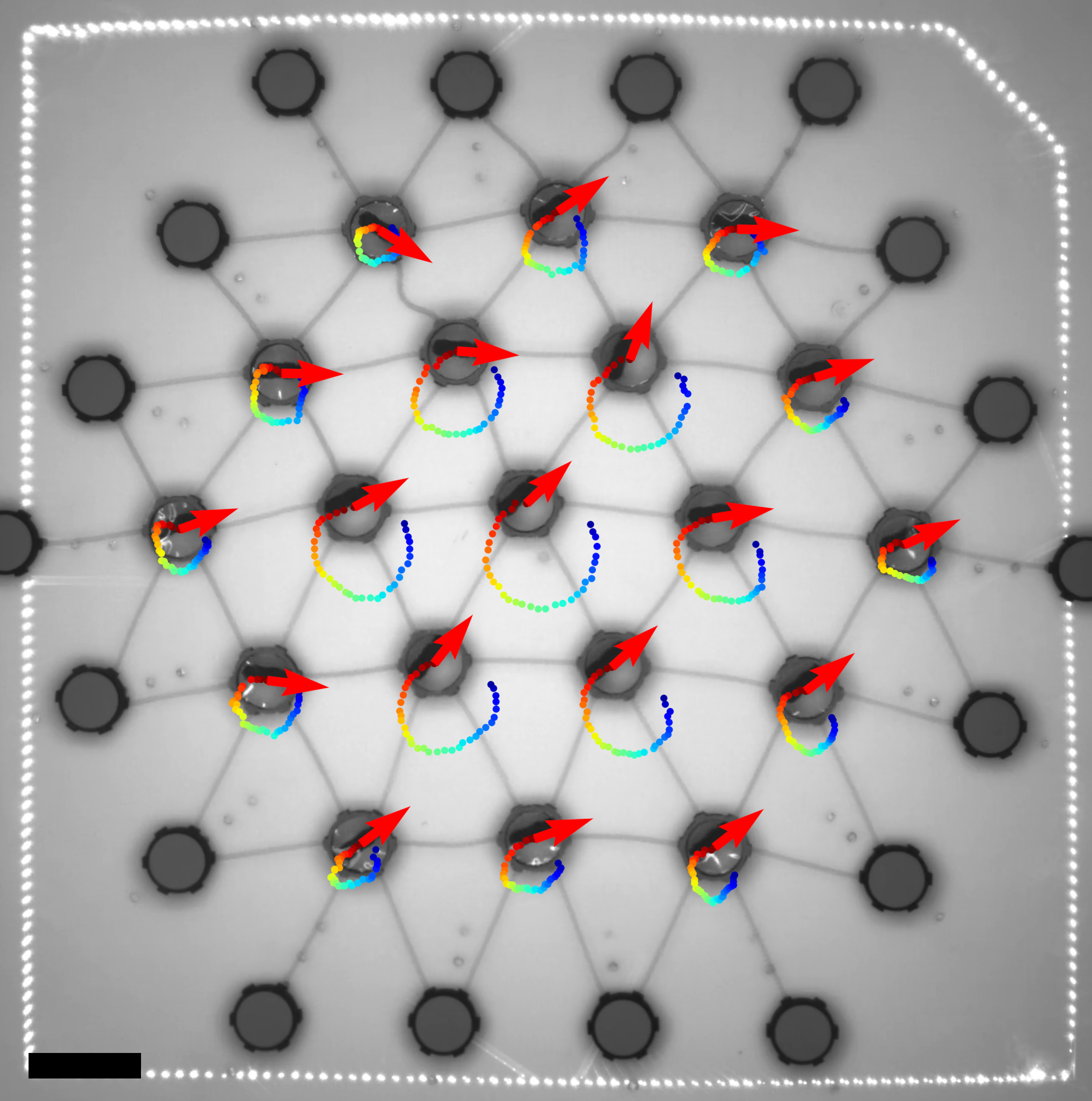}};
\node[anchor=south west] at (7.75,7.1)
{\includegraphics[width=2.0cm]{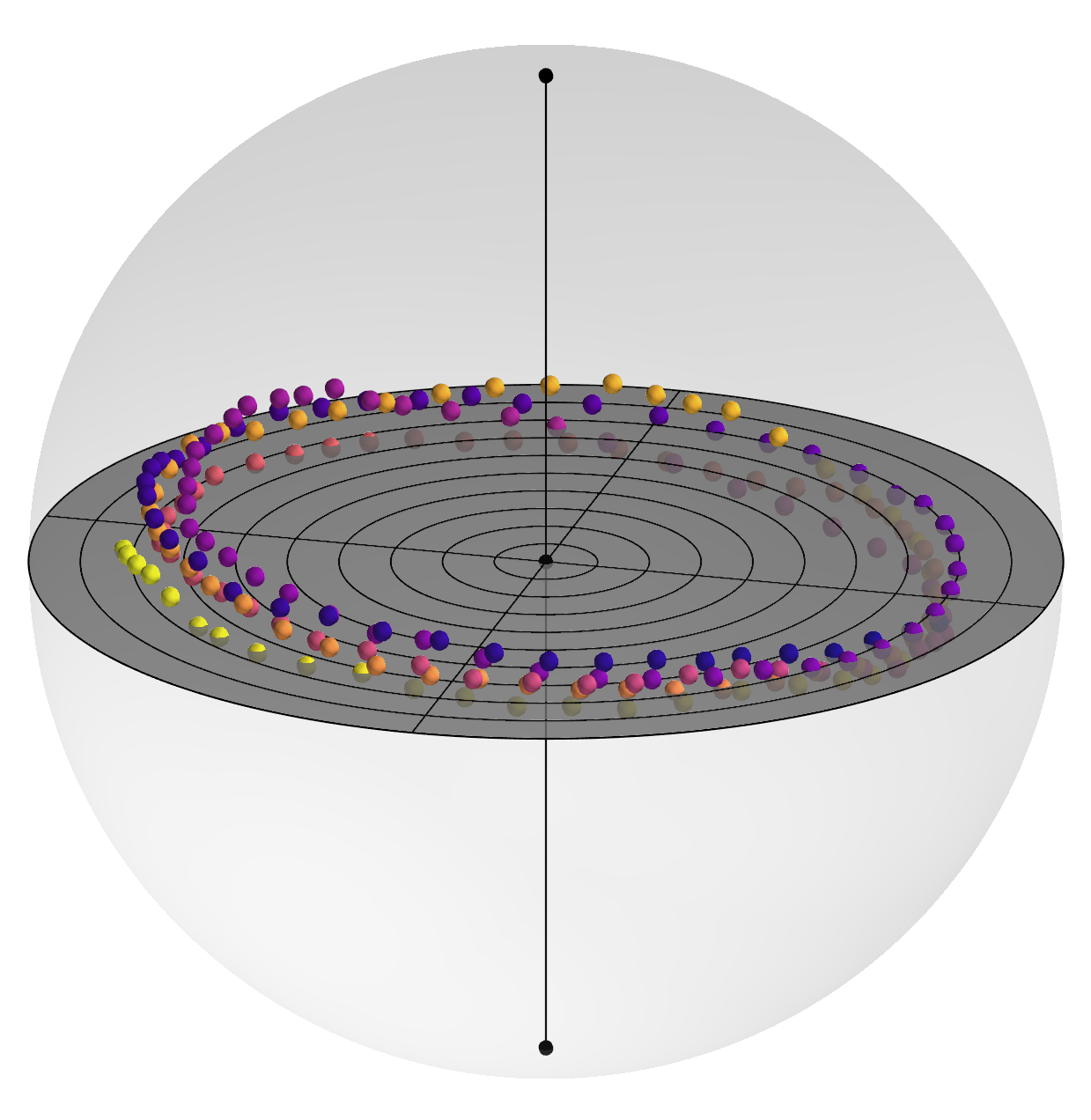}};

\node[] at (1.55,9.0) {\small (a)};
\node[] at (5.95,9.0) {\small (b)};

\node[] at (1.75,6.65) {\small (c)};
\node[] at (6.0,6.65) {\small (d)};

\node[] at (1.45,3.48) {\small (e)};
\node[] at (5.83,3.48) {\small (f)};

\end{tikzpicture}
\vspace*{-0.7cm}
\caption{\small{\textbf{Experimental realization of a controlled switch between collective actuations}. (a) GAR in a triangular lattice under central pinning; $N = 36$. (b) SLO in a triangular lattice under edge pinning; $N = 19$. 
The left panel displays the dynamics in real space (red arrows: polarities $\boldsymbol{\hat{n}}_i$; trajectories color coded from blue to red with increasing time; scale bars: $10$ cm); the right panel displays the dynamics projected on the translation and rotation modes of the structures (vertical axis: $\langle R | \boldsymbol{\hat{n}} \rangle$, equatorial plane: $\langle T_{x/y} | \boldsymbol{\hat{n}} \rangle$), see notations and convention in panel (c).
(d) A switch between GAR and SLO is obtained in a model elastic structure with fixed pinning conditions by tuning tension: condensation fraction on the rotation mode $\lambda_{R}$ as a function of tension $\alpha - 1$  (red bullets are obtained from data averaged in the steady state; the gray dots and arrows sketch the transitory regime when initials conditions enforce rotation at large tension). (e-f) GAR, resp. SLO in the active \textit{Gerris} at low ($\alpha = 1.0$), resp. large ($\alpha = 1.8$) tension; $N=6$.}}
\label{fig:experiments}
\end{figure}

At first, autonomous actuation was demonstrated in the case of active elastic networks comprising zero modes mechanisms~\citep{ferrante2013elasticity, woodhouse2018autonomous}. More recently, it was shown both experimentally and numerically that a mechanically stable elastic structure can also exhibit collective actuation (CA), when a nonlinear elasto-active feedback is present~\citep{baconnier2022selective}. Typical examples of such CA are illustrated on Fig.~\ref{fig:experiments}. When a given node of the elastic structure is pinned both in translation and rotation, the structure alternatively rotates clockwise and counter-clockwise around this node (Fig~\ref{fig:experiments}-a and SM Movie 1). When the structure is pinned at its edge, the nodes perform a local but synchronized oscillation that spontaneously breaks chiral symmetry (Fig~\ref{fig:experiments}-b and SM Movie 2). In the following, we shall call GAR (Global Alternating Rotation), resp. SLO (Synchronized Local Oscillations), these two regimes. Quite remarkably, it was shown that the CA obeys a non trivial selection mechanism in the sense that it does not necessarily correspond to a condensation on the lowest energy modes of the elastic structure~\citep{baconnier2022selective}. In the above example the two CA regimes are obtained by imposing different pinning conditions. This is not always a convenient way of monitoring the actuation of a structure and, certainly, there are circumstances under which the use of a continuous control parameter is desirable.

In this letter, we demonstrate that mechanical tension is a suitable mechanism for controlling CA in active solids. Tension governs the sound of stringed musical instruments, the growth and response of biological systems \cite{bunting1926effect, kolega1986effects, hinz2001mechanical, ingber2006cellular, anava2009regulative}, and the stability of civil engineering work \cite{zhang2015tensegrity}. We show how it can be harnessed to manipulate the vibrational spectrum of 2$d$ active solids, and control the switch between different CA regimes. We establish the experimental proof of concept using a toy-model active solid (Figs.~\ref{fig:experiments}-(e,f) and SI Movies 3-4), dissect the underlying mechanism and extend our findings to more general geometries on the basis of the numerical study of an agent based model and theoretical arguments. 
%We find that the selectivity of CA, allows for a control, that goes beyond the adjustment of the energies of the elastic modes.

Our prototypic active solids, described in detail in~\cite{baconnier2022selective}, consist in elastic structures, composed of $N$ active units connected by springs of stiffness $k$ and rest length $l_0$ (see Fig~\ref{fig:experiments}). Each active unit is made of a Hexbug\copyright, a centimetric battery powered running robot, embedded in a 3d printed cylinder (height $1.4$ cm; internal radius $2.50$ cm). Each active unit exerts a polar force $F_0 \boldsymbol{\hat{n}}_i$, where $\boldsymbol{\hat{n}}_i$ denotes the orientation of the Hexbug.  The dynamics of the active elastic structures are captured and tracked at 40 frames per second. Each node has a well-defined reference position, but is displaced by the active unit. In contrast, the polarity of each unit is free to rotate and reorients towards the node's displacement rate $\dot{\boldsymbol{u}}_i$. This nonlinear elasto-active feedback between deformations and polarities is controlled by the ratio $\pi = l_e/l_a$ with $l_e = F_0/k$, the typical elastic deformation caused by the active force and $l_a$, the alignment length over which $\boldsymbol{\hat{n}}_i$ aligns towards $\dot{\boldsymbol{u}}_i$. 

\begin{figure*}[t]
%\captionsetup{format=plain}
\centering
\hspace*{-6.6mm}
\begin{tikzpicture}
   
\node[rotate=90] at (2.1,0.35)
{\includegraphics[height=3.5cm]{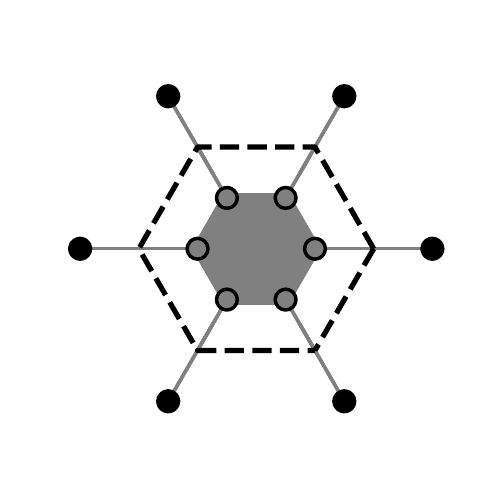}};

\draw[<->] (2.28,0.80) -- (2.28,1.65);
\node[] at (2.55,1.3) {\footnotesize $\alpha l_0$};

\node[rotate=30] at (1.65,1.18) {\footnotesize $\alpha = 1$};

\node[] at (4.9,1.6)
{\includegraphics[width=0.9cm]{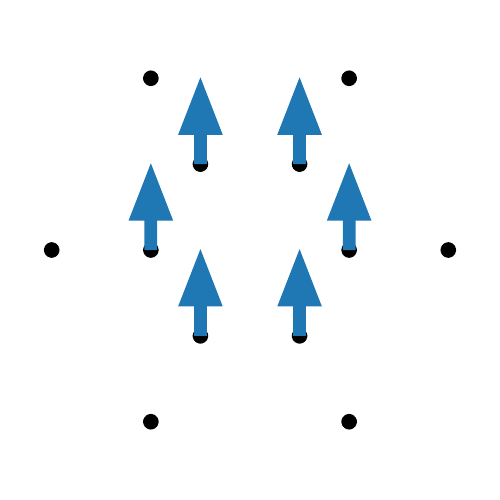}};
\node[] at (5.8,1.6)
{\includegraphics[width=0.9cm]{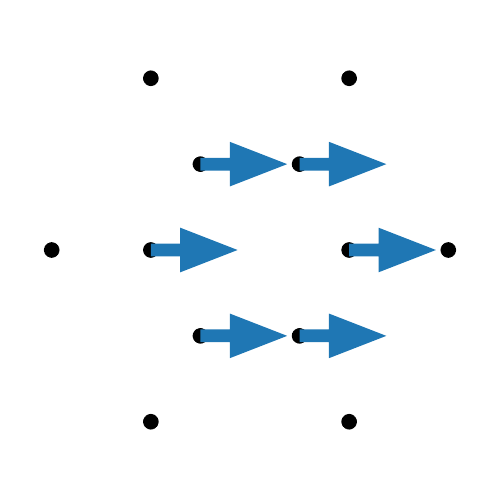}};
\node[] at (6.97,-0.83)
{\includegraphics[width=0.9cm]{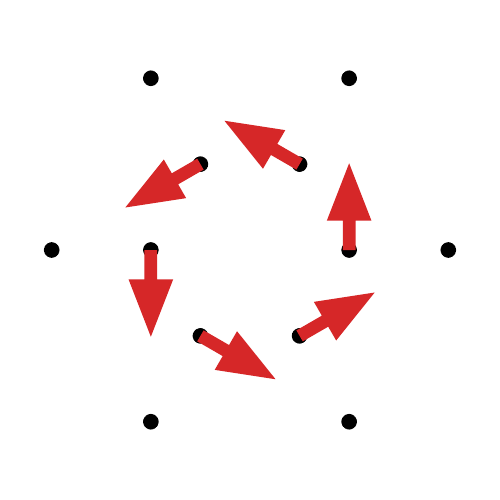}};

\node[rotate=15] at (7.07,1.05) {\small $\color{color2} | R \rangle$};
\node[rotate=7] at (7.07,1.65) {\small $\color{color1} | T \rangle$};

\node[] at (5.5,0.0)
{\includegraphics[height=5.0cm]{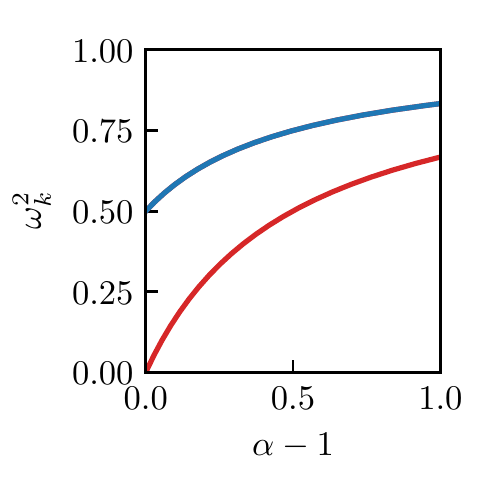}};

\node[] at (10.0,0.07)
{\includegraphics[height=5.2cm]{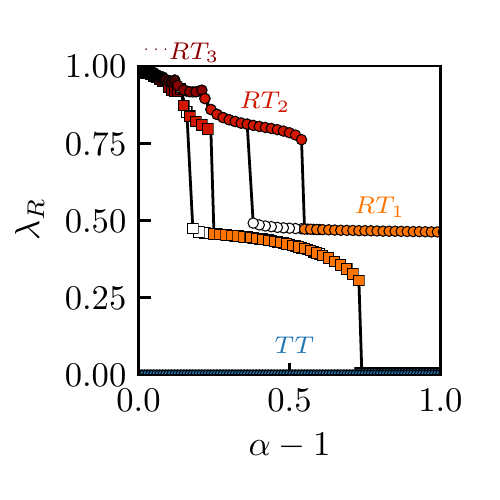}};

\draw[->] (9.91,0.6) -- (9.88,1.1);
\draw[->] (9.27,0.6) -- (9.24,1.1);

\draw[->] (10.66,1.1) -- (10.67,0.6);
\draw[->] (9.71,1.1) -- (9.72,0.6);

\draw[->] (11.25,-0.45) -- (11.26,-0.95);

\draw[<->] (9.1,-1.05) -- (11.7,-1.05);
%\node[] at (10.0,0.0)
%{\includegraphics[height=5.0cm]{img/annealing/alphaAnnealing_Linear_NLinear.pdf}};

%\node[] at (13.5,0.0)
%{\includegraphics[height=2.5cm]{img/annealing/regimes/40.png}};
\node[] at (13.3,1.0)
{\includegraphics[height=2.0cm]{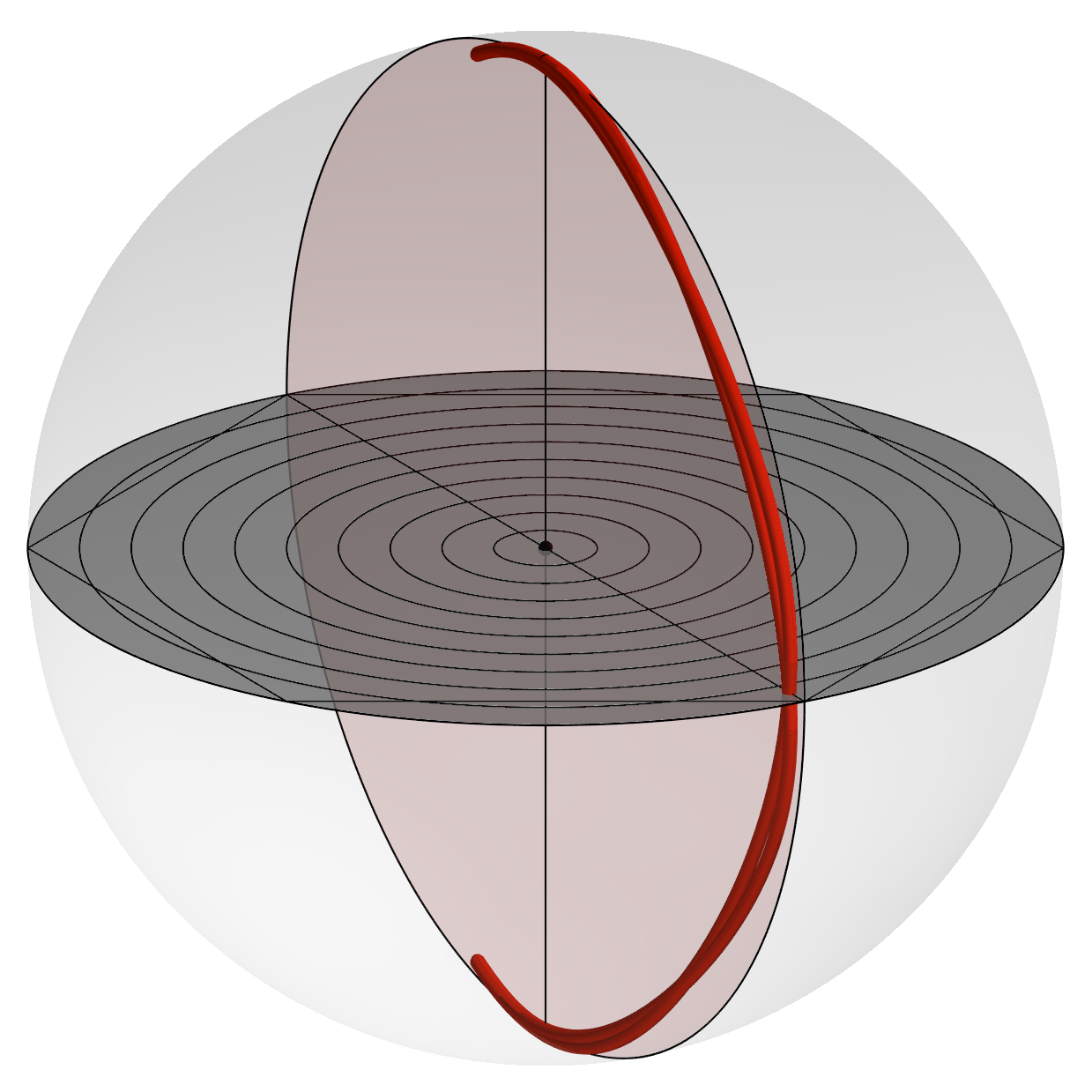}};
\node[] at (15.4,1.0)
{\includegraphics[height=2.0cm]{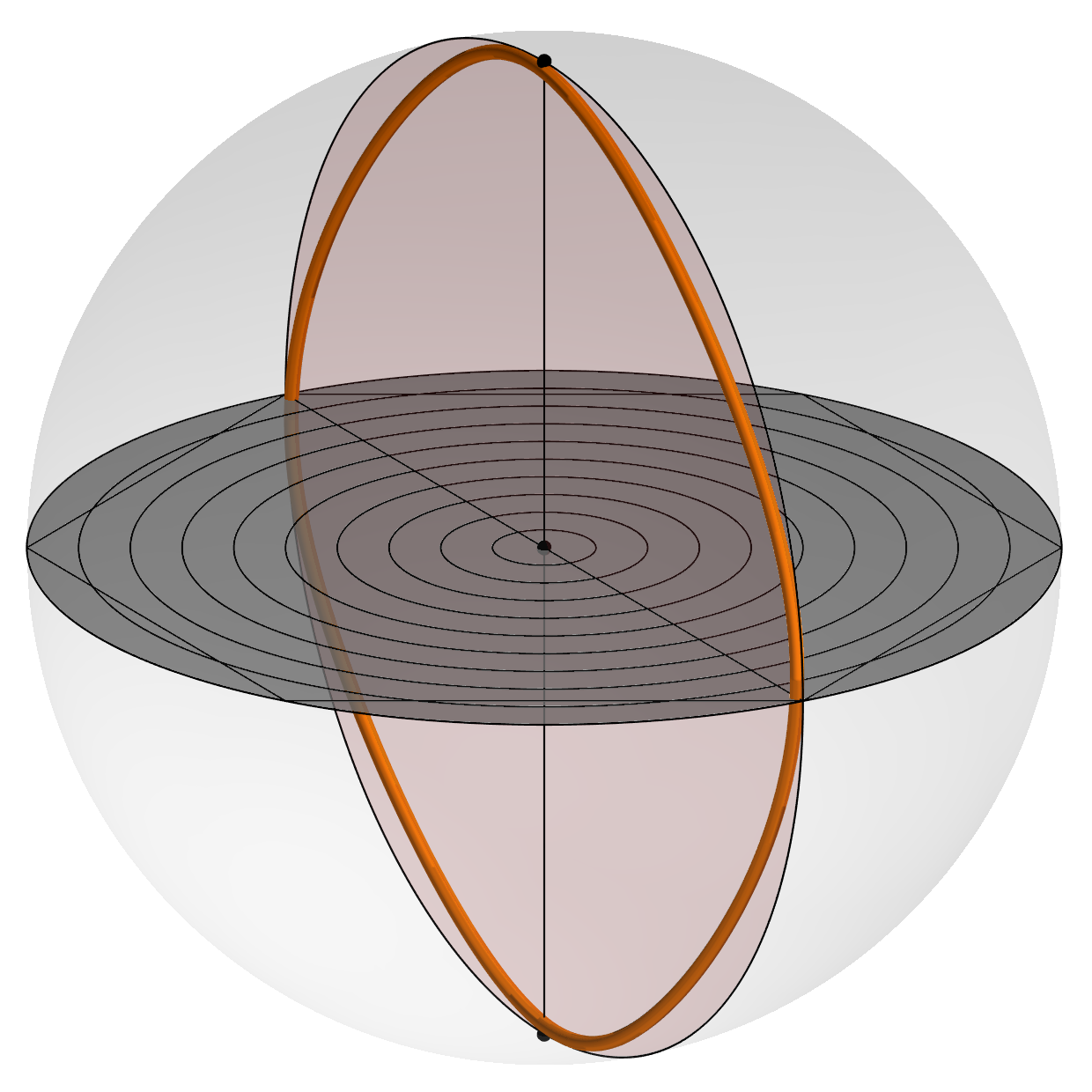}};
\node[] at (17.5,1.0)
{\includegraphics[height=2.0cm]{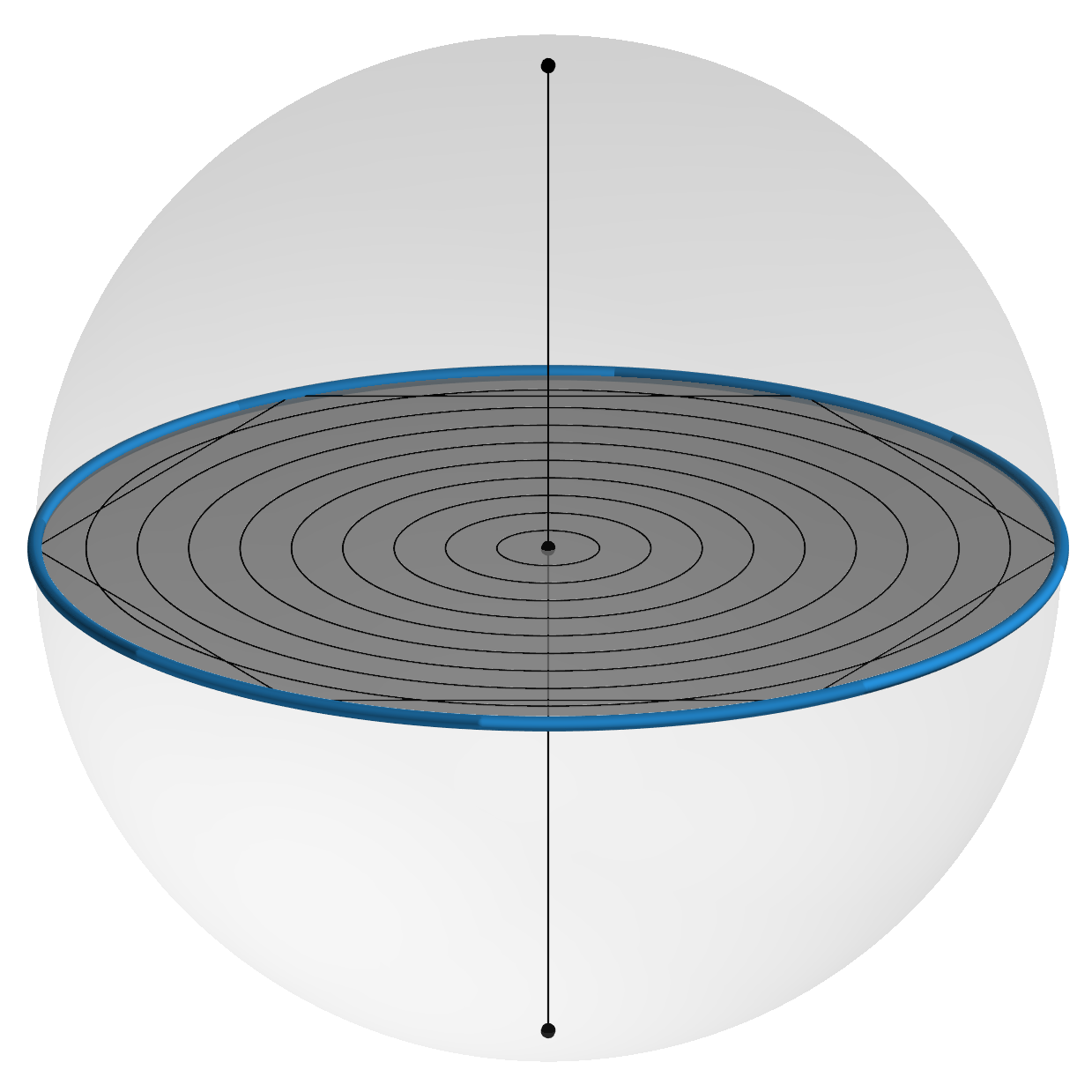}};

\node[] at (13.3,-1.0)
{\includegraphics[height=2.0cm]{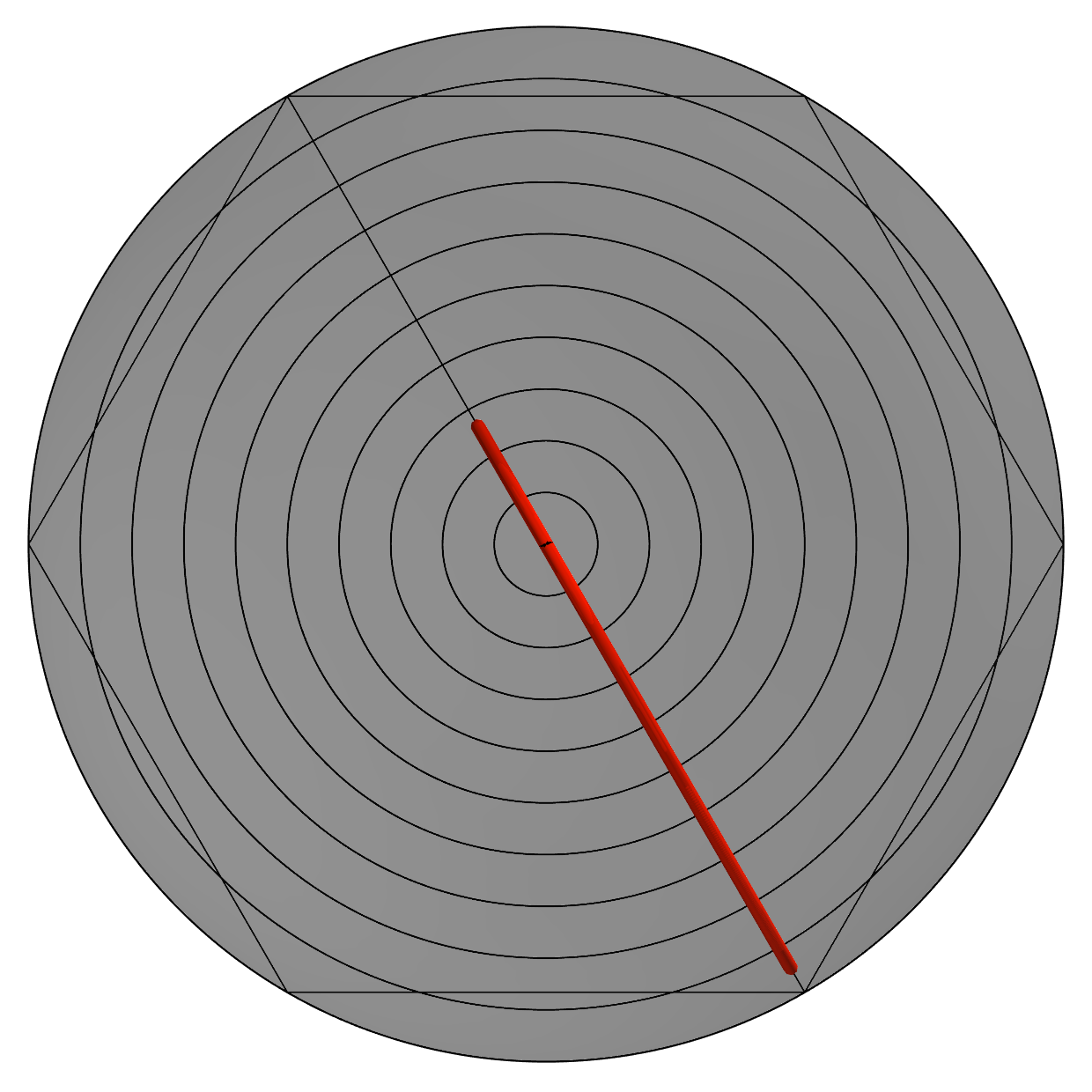}};
\node[] at (15.4,-1.0)
{\includegraphics[height=2.0cm]{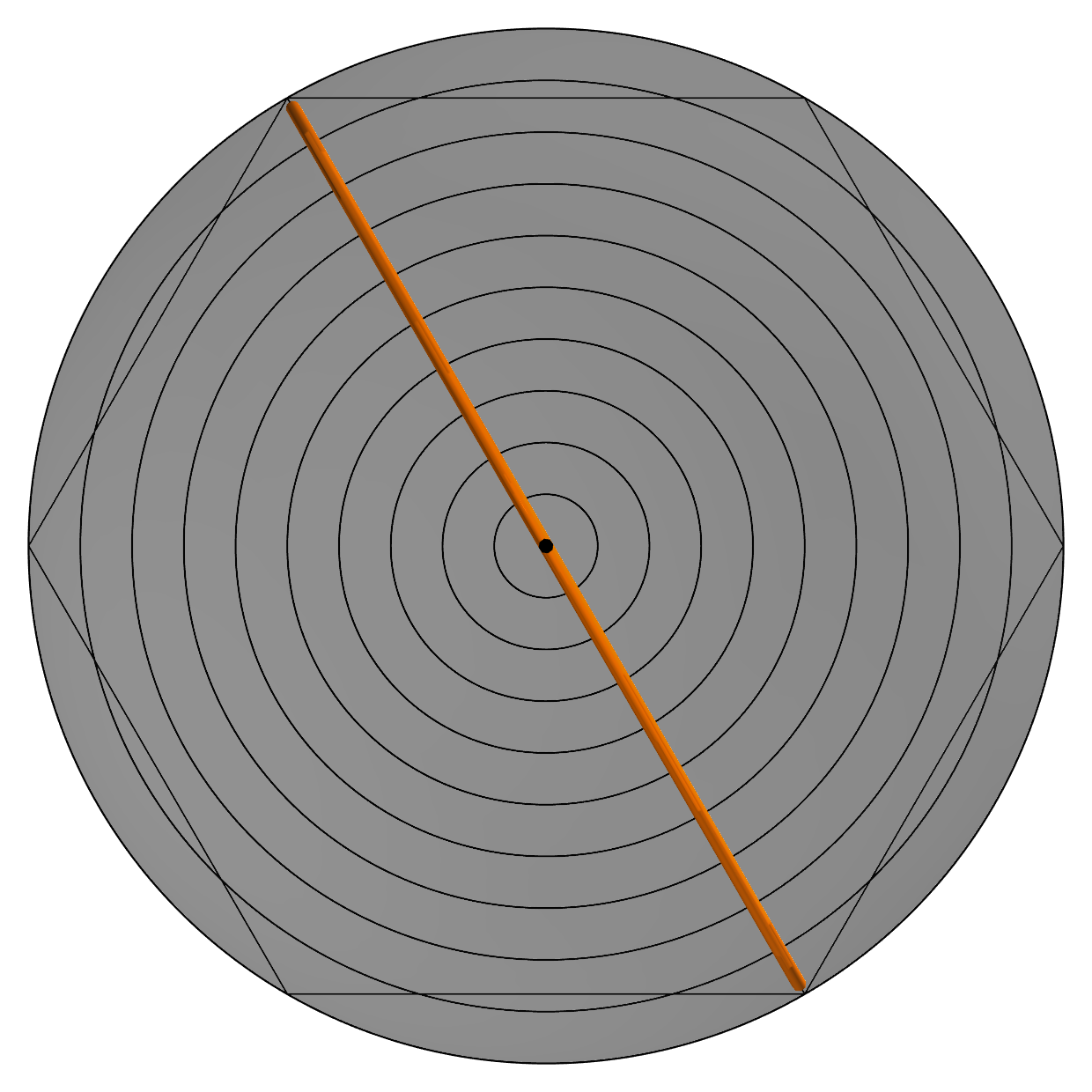}};
\node[] at (17.5,-1.0)
{\includegraphics[height=2.0cm]{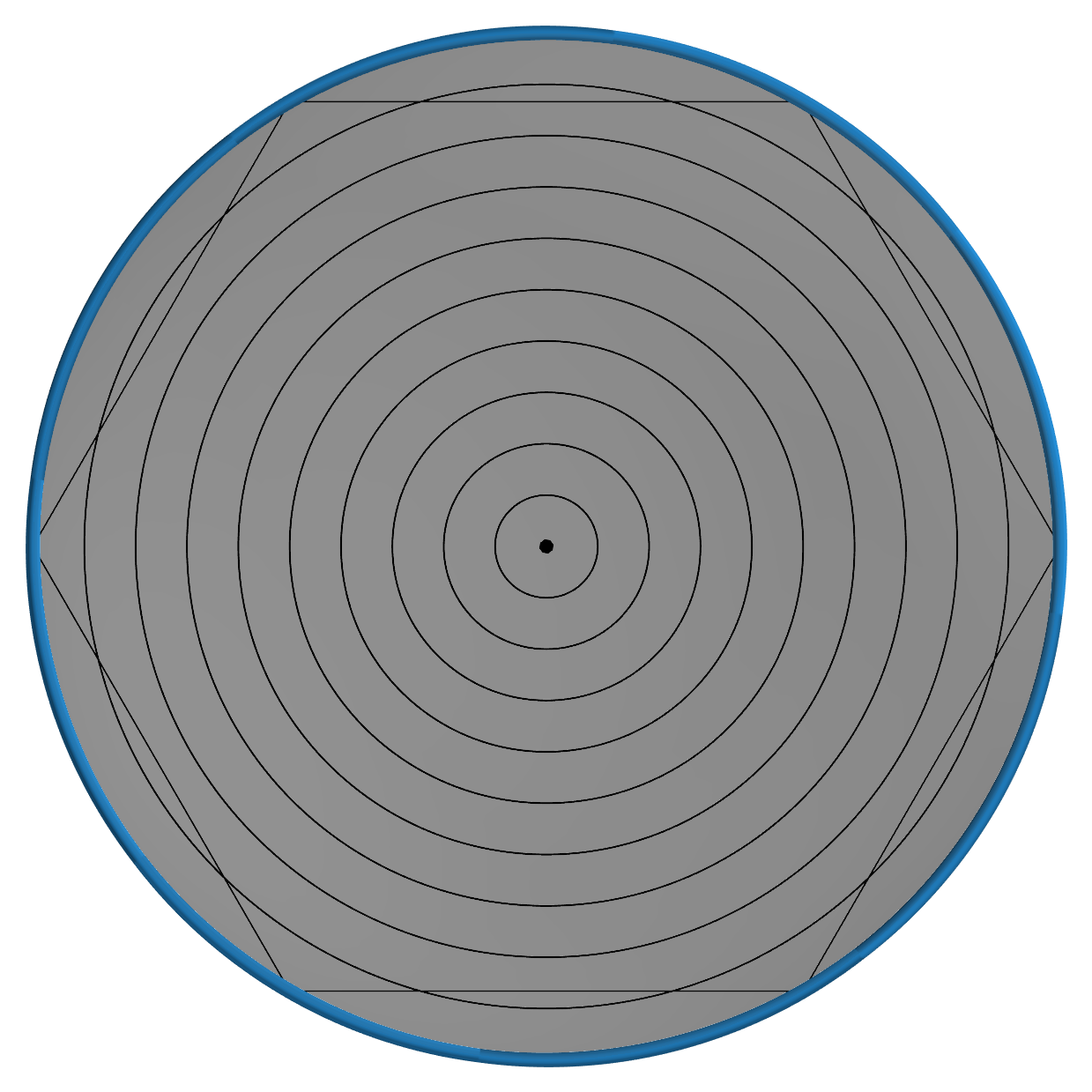}};

\draw[->] (16.12,1.8) arc (110:70:1.0);
\draw[<->] (14.02,1.8) arc (110:70:1.0);

\node[] at (12.65,-0.0) {\small \color{color_RT2} $RT_2$};
\node[] at (14.75,-0.0) {\small \color{color_RT1} $RT_1$};
\node[] at (16.85,-0.0) {\small \color{color1} $TT$};

\node[] at (13.85,-0.0) {\small x$6$};
\node[] at (15.95,-0.0) {\small x$6$};
\node[] at (18.05,-0.0) {\small x$2$};

\node[] at (2.1,2.4) {\small (a)};
\node[] at (5.9,2.4) {\small (b)};
\node[] at (10.4,2.4) {\small (c)};

%\node[] at (13.3,2.4) {\small (d)};
\node[] at (15.4,2.4) {\small (d)};
%\node[] at (17.5,2.4) {\small (f)};

\end{tikzpicture}
\vspace*{-1.2cm}
\caption{\small{\textbf{Active \textit{Gerris}'s dynamics as a function of lattice tension.} (a) Elastic architecture cartoon. The gray area illustrates the rigid inner hexagon. (b) Normal modes spectrum as a function of springs tension $\alpha - 1$. The red (resp. blue) solid line corresponds to the rotation mode (resp. degenerated translational modes). (c) Condensation fraction on the rotation mode $\lambda_{R}$ as a function of lattice tension $\alpha - 1$ (yellow-orange to black symbols : the different $RT$ regimes; blue symbols $TT$ regimes; ($\circ$): harmonic approximation; (\scalebox{0.6}{$\square$}): including geometrical non-linearities; empty markers: backward annealing). (d) Side and top view of the $3d$ representations of the polarity field steady dynamics projected on the rotation and translation modes for the regimes $RT_2$ (left); $RT_1$ (middle) and $TT$ (right) (vertical axis: $\langle R | \boldsymbol{\hat{n}} \rangle$, equatorial plane: $\langle T_{x/y} | \boldsymbol{\hat{n}} \rangle$).}}
\label{fig:design_principle}
\end{figure*}
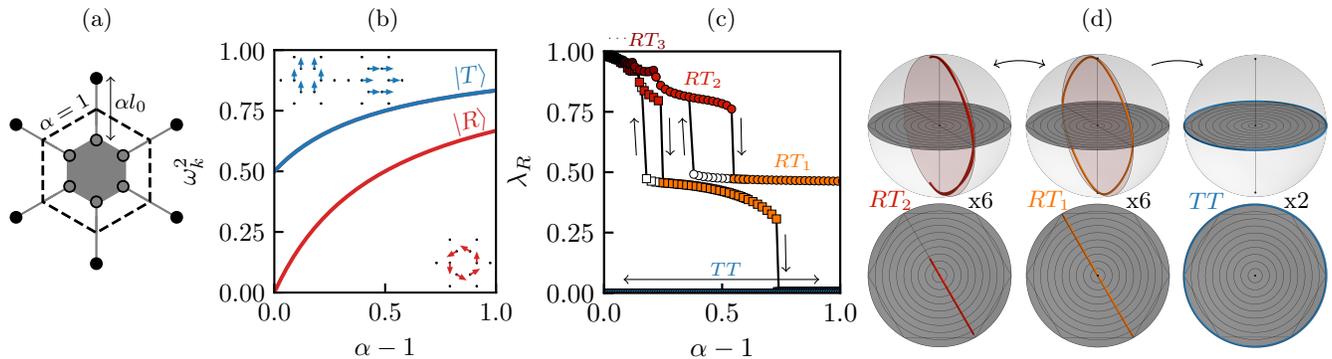

We start by demonstrating experimentally the possibility of controlling the switch between the two CA regimes described above and so far obtained with different pinning conditions. To do so, we design an active elastic structure, which consists of $N=6$ active particles at the vertices of an inner rigid hexagon, each connected radially to the vertices of an outer pinned hexagon via soft springs of stiffness $k = 1$ N/m (Fig. \ref{fig:design_principle}-a), which we call active \textit{Gerris}. 
We control the tension in the springs by elongating homogeneously the radial springs of a factor $\alpha$.  As illustrated in Figs.~\ref{fig:experiments}-(d-f), the dynamics of the active \textit{Gerris} switches from the GAR regime (Fig. \ref{fig:experiments}-e and SI Movie 3) at low tension, to the SLO regime (Fig. \ref{fig:experiments}-f and SI Movie 4) at large tension.

These dynamics are best described when decomposed on the elastic modes of the structure, that are the eigenvectors, $| \boldsymbol{\varphi}_k \rangle$, associated with the eigenvalues,  $\omega_{k}^{2}$, of the dynamical matrix, $\mathbb{M}$. More specifically, we shall represent the dynamics in the space spanned by the amplitude of the polarity field projected on the three modes of interest. By convention, the vertical axis represents the normalized projection on the rotation mode $a_{R} = \langle R | \boldsymbol{\hat{n}} \rangle / \sqrt{N}$, whereas the equatorial plane represents the normalized projections on the two translational like modes $a_{T_{x/y}} = \langle T_{x/y} | \boldsymbol{\hat{n}} \rangle /\sqrt{N}$ (Fig \ref{fig:experiments}-c). From the polarity field normalization, the projections are confined inside the $3$-sphere of radius $\sqrt{N}$, normalized to $1$. 
In the GAR regime, obtained from the central pinning condition, the dynamics alternatively condensate on the clockwise and counterclockwise rotation (the poles of the sphere), separated by fast reversal motion (Fig~\ref{fig:experiments}-a). In the SLO regime, obtained from the edge pinning condition, the dynamics condensate on the translational modes spanning the equator of the sphere (Fig.~\ref{fig:experiments}-b). Figs.~\ref{fig:experiments}-e and f convincingly demonstrate that the active \textit{Gerris} explores the same dynamics under the control of tension. The dynamics are quantified by computing the mean square projection of the polarity field on each mode:
\begin{equation}
 \lambda_{k} = \langle a_k^2 \rangle_t = \frac{1}{T} \int^{T} \left[ \frac{\langle \boldsymbol{\varphi}_{k} | \boldsymbol{\hat{n}} (t) \rangle}{\sqrt{N}} \right]^{2} dt
\end{equation}
The active \textit{Gerris} switch is illustrated by the abrupt drop of this condensation fraction on the rotation mode $\lambda_{R}$ as tension increases (Fig. \ref{fig:experiments}-d). 

To investigate the origin of the switch in the active \textit{Gerris}, we now proceed to numerical simulations using an agent based model, which was shown to faithfully describe the dynamics of active elastic structures~\citep{baconnier2022selective}:
\begin{subequations} 
\vspace{-0.2cm}
\label{eq:dimensionless_noiseless}
\renewcommand{\theequation}{\theparentequation.\arabic{equation}}
\begin{align}
 \dot{\boldsymbol{u}}_{i} &= \pi \boldsymbol{\hat{n}}_{i} + \boldsymbol{F}_{i}^{el}, \label{eq1:dimensionless_noiseless} \\
 \dot{\boldsymbol{n}}_{i} &=  (\boldsymbol{\hat{n}}_{i} \times \dot{\boldsymbol{u}}_i ) \times \boldsymbol{\hat{n}}_{i} + \sqrt{2D}\xi_i \boldsymbol{\hat{n}}_i^{\perp},
 \label{eq2:dimensionless_noiseless} 
\end{align}
\end{subequations}
where $\boldsymbol{u}_i$ is the displacement of node $i$ with respect to its reference position, and $\xi_i$ are i.i.d gaussian variables with zero mean and correlations $\langle \xi_i(t) \xi_j (t') \rangle = \delta_{ij} \delta (t - t')$. The elasto-active feedback, $\pi$ controls the emergence of CA. We set it to $\pi = 2.0$, a value consistent with the experiments, and investigate the effect of tension.

The \textit{Gerris} has six nodes that are connected by a structure, which can safely be considered as rigid (Fig.~\ref{fig:design_principle}-a and~\citep{Supplemental_information}). It is thus described  by three degrees of freedom, the spatial coordinates of its barycenter and its angular orientation, the dynamical equations of which are provided in~\citep{Supplemental_information}. The three associated normal modes are two degenerated translation modes $|T_{x/y}\rangle$ and one rotation mode $|R\rangle$, which are illustrated on Fig.~\ref{fig:design_principle}-b, together with their energies as a function of the imposed tension. Both the rotation and translation energies increase with tension, but the energetic ordering of the modes is preserved, and their geometries are unaffected. The three modes end up degenerated at infinite tension. 

We first simulate the noiseless, $D=0$, active \textit{Gerris} equations in the harmonic approximation ($\boldsymbol{F}_{i}^{el} = - \mathbb{M}_{ij} \boldsymbol{u}_j$), annealing back and forth between small and large tensions. We find two linearly stable actuation branches, which we respectively denote the $TT$ and $RT$ regimes (Fig. \ref{fig:design_principle}-c, circle markers). 
The $TT$ regime is a strict condensation of the polarity field on the equator (Fig.~\ref{fig:design_principle}-d), with $\lambda_{R}=0$, corresponding to a SLO of the \textit{Gerris}. This regime exactly maps to that of a single particle trapped in a parabolic potential \citep{dauchot2019dynamics, baconnier2022selective, Supplemental_information}. The $RT$ regimes consist in a condensation of the polarity field on a plane, defined by the rotation vector $|R\rangle$ and one of the six translational vector $|T\rangle$, pointing toward one of the hexagon's main axis, in the equatorial plane (Fig.~\ref{fig:design_principle}-d).  They correspond to a GAR of the \textit{Gerris}. The six possible orientations of this plane, define six equivalent attractors, one of which is selected, spontaneously breaking the $6$-fold symmetry of the system~\citep{Supplemental_information}.  Depending on the tension, we actually report different $RT$ dynamics, separated by hysteretic transitions, which differ in the precise trajectory of the alternating rotation. These RT regimes and the transitions among them are well captured by the dynamics of a single particle trapped in an elliptic harmonic potentials~\cite{Supplemental_information, damascena2022coexisting}. Within the linear level of description, there is however no switch between the coexisting $RT$ and $TT$ regimes.

Including the geometrical non-linearities of the elastic forces, allows for a better description of the experimental observations (Fig. \ref{fig:design_principle}-c, square markers). Simulating the full expression for central force springs, we find that the $TT$ regime is unaffected, while the stability range of the $RT$ regimes are shifted toward smaller tensions. More significantly, the $RT$ regime destabilizes towards the $TT$ regime for large enough tension. Therefore, geometrical non-linearities allow for a switch between the $RT$ and the $TT$ regimes as tension increases. 

\begin{figure}[t!]
%\captionsetup{format=plain}
\centering
\hspace*{0.0mm}
\begin{tikzpicture}

\node[] at (0.85,0.0)
{\includegraphics[height=2.2cm]{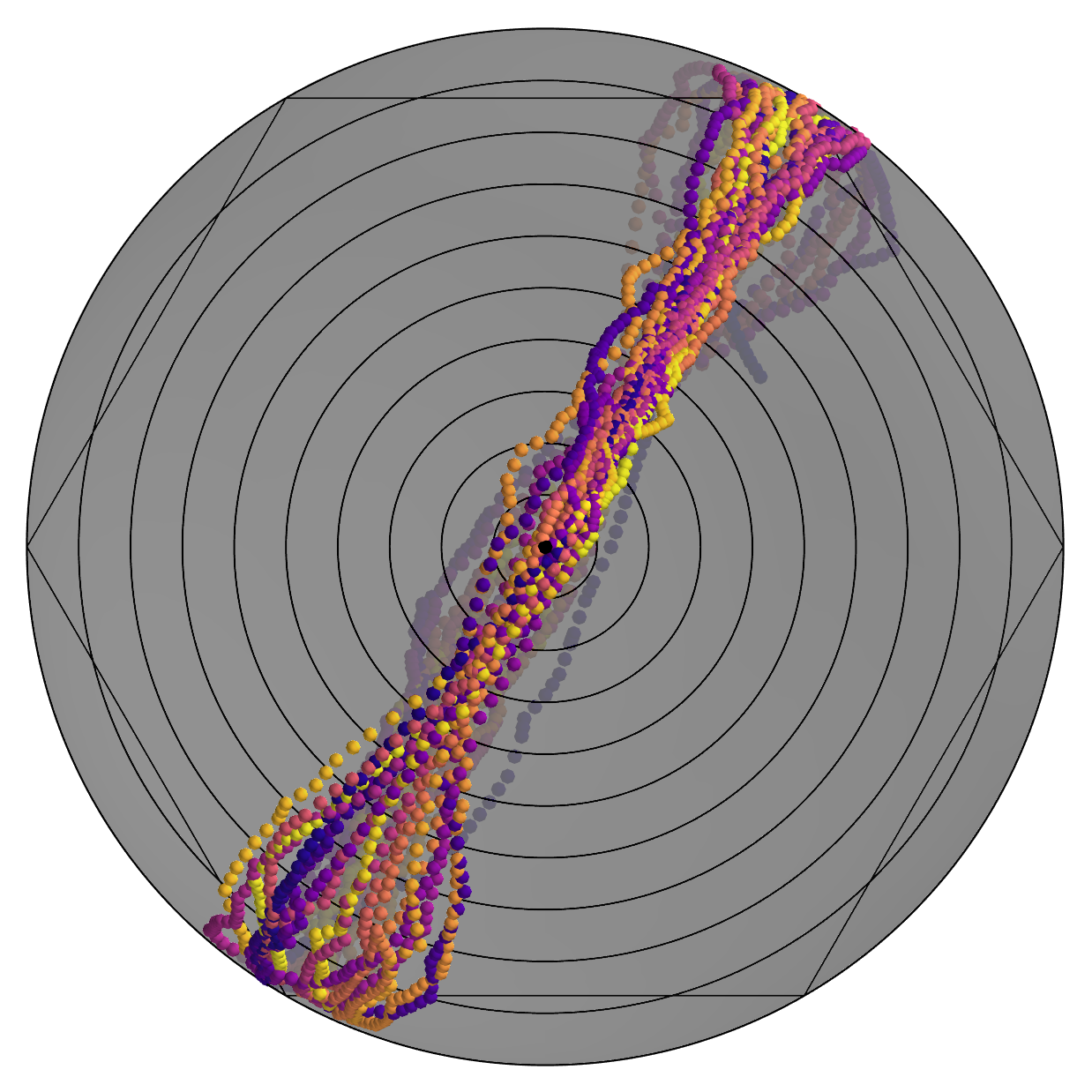}};
\node[] at (3.1,0.0)
{\includegraphics[height=2.2cm]{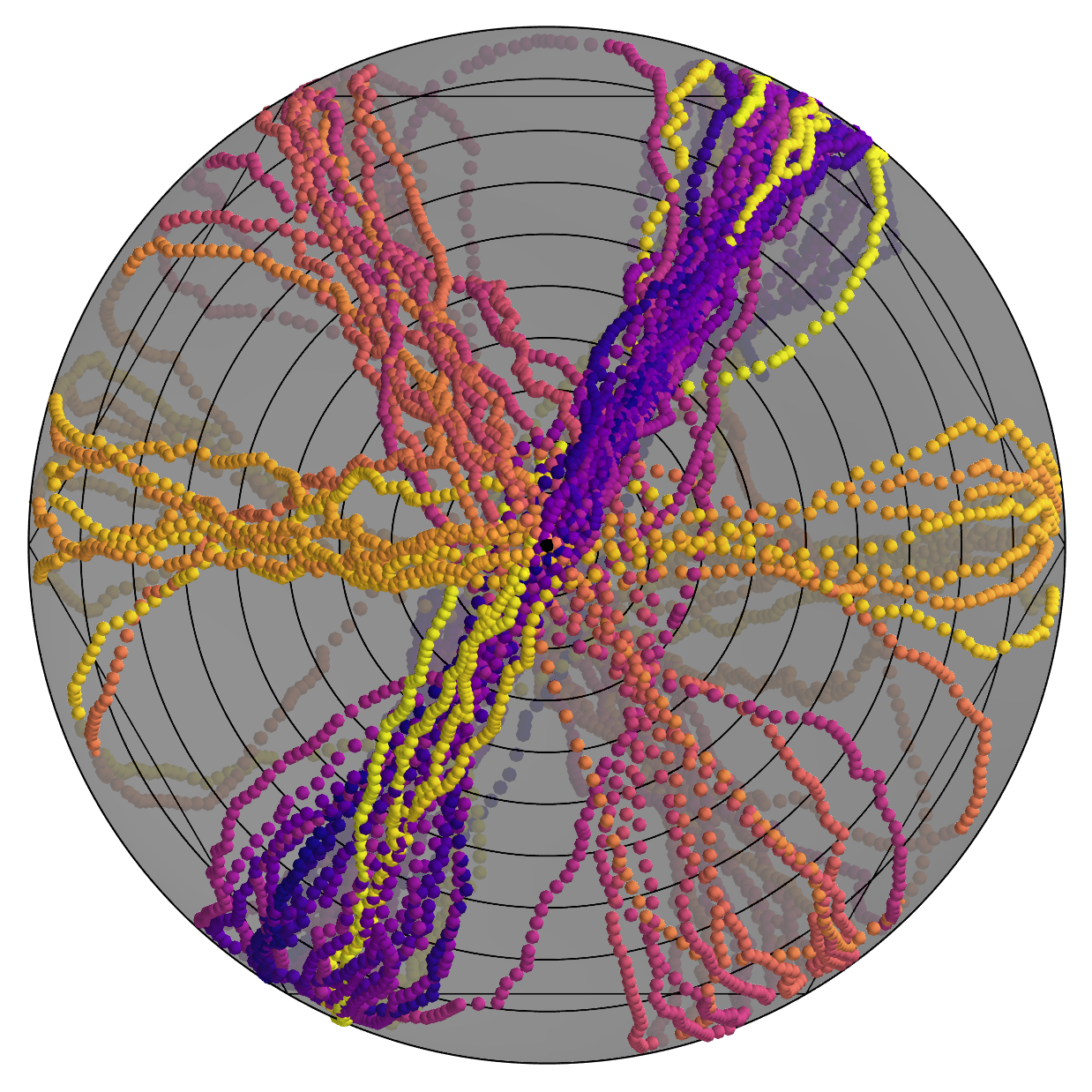}};
\node[] at (5.35,0.0)
{\includegraphics[height=2.2cm]{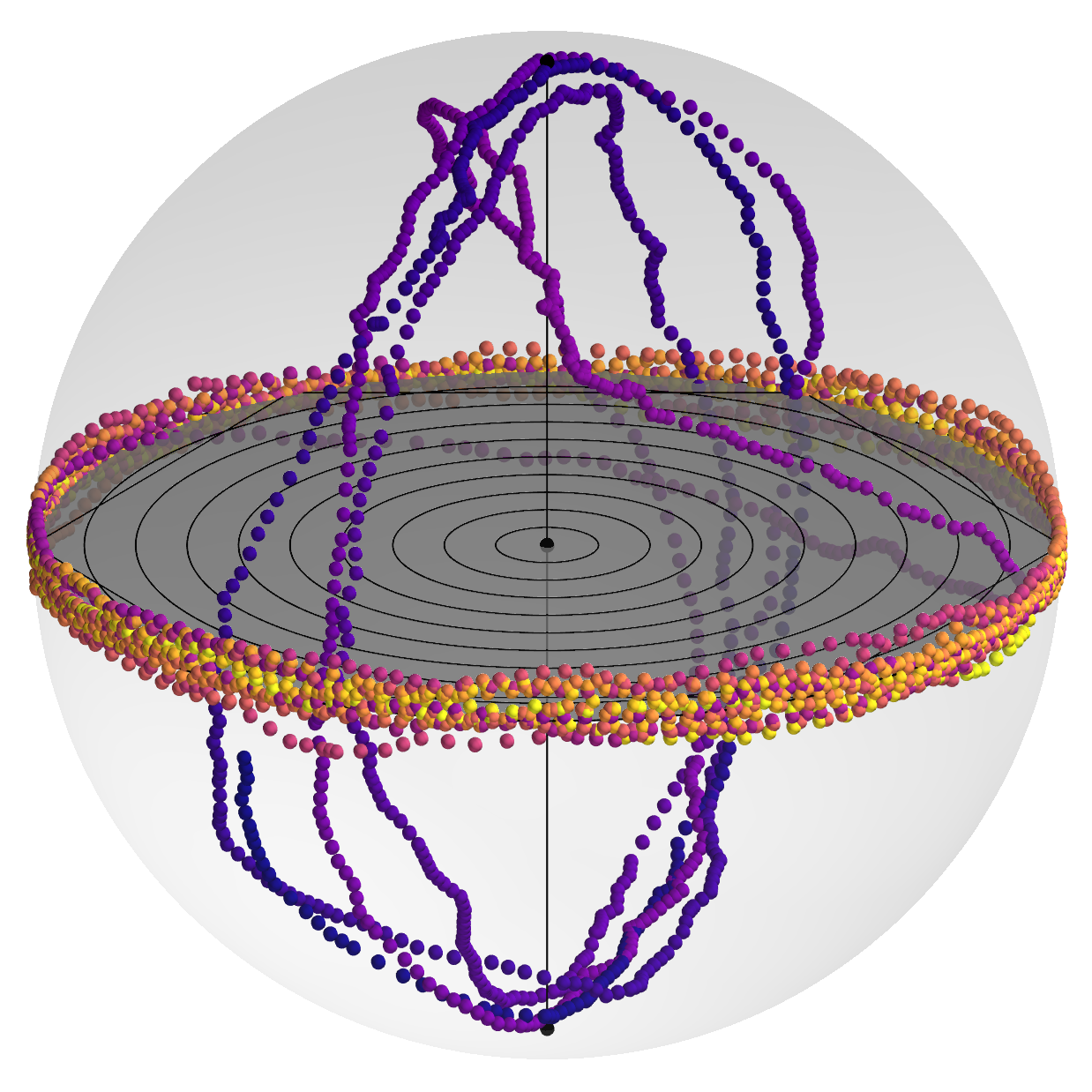}};

\node[] at (0.85,1.3) {\small $\tau_{\text{short}}$};
\node[] at (3.1,1.3) {\small $\tau_{\text{inter}}$};
\node[] at (5.35,1.3) {\small $\tau_{\text{long}}$};

\node[] at (3.5,-3.8)
{\includegraphics[width=7.0cm]{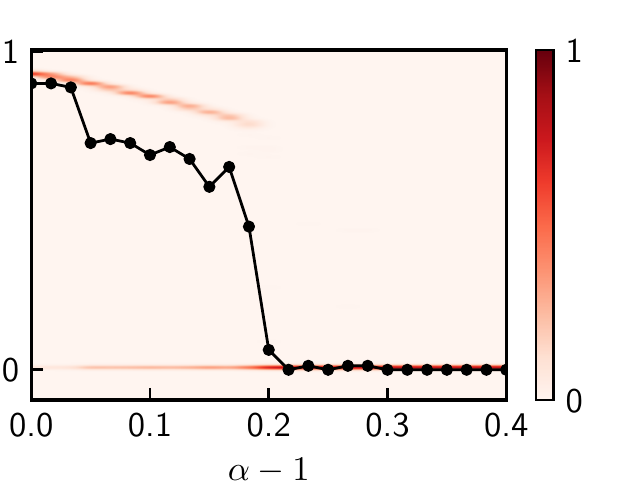}};

\node[rotate=90] at (-0.2,-3.4) {\normalsize $\rho_{\lambda_{R}}, \mathbb{P}_{RT}$};

\node[] at (-0.5,1.0) {\small (a)};
\node[] at (5.35,-1.85) {\small (b)};

\end{tikzpicture}
\vspace*{-0.5cm}
\caption{\small{\textbf{The active \textit{Gerris}'s dynamics in the presence of noise} (a) Projection of the polarity field dynamics illustrating the effect of noise on the $RT_1$ regime, at short, intermediate, and long times (vertical axis: $\langle R | \boldsymbol{\hat{n}} \rangle$, equatorial plane: $\langle T_{x/y} | \boldsymbol{\hat{n}} \rangle$). (b) Density of condensation fraction on the rotation mode $\rho_{\lambda_{R}}$ (color coded as given by the colorbar) and probability to end in a $RT$ regime at long times $\mathbb{P}_{RT}$ (black markers) as a function of tension $\alpha - 1$.}}
\label{fig:non_linear_noisy}
\end{figure}
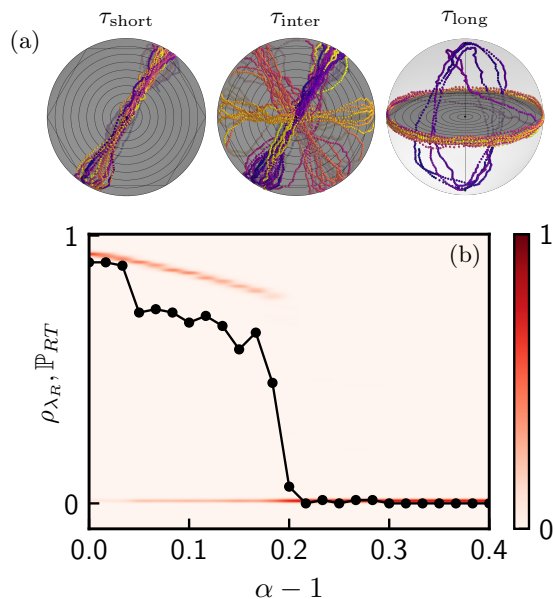

The $TT$ regime persists for all values of the tension and coexists with the $RT$ regime. This raises the issue of  the relative stability of the two attractors. We address it by adding a small noise, $D=10^{-2}$, consistent with the experimental values. Starting from the $RT$ regime, the system first remains close to the initial $RT$ attractor, then visits the six equivalent $RT$ attractors, before it eventually undergoes a destabilization event, which drives it into the $TT$ regime at long times (Fig. \ref{fig:non_linear_noisy}-a). The smaller the tension, the longer it takes for this destabilization event to take place. We evaluate the metastability of the $RT$ regime, by performing $80$ independent simulation runs with random initial condition, for each value of the tension. At small tension, the probability to end up in a $RT$ regime at $t=10000$, $\mathbb{P}_{RT}$, is close to one and slowly decreases with increasing tension. This is due to both the increasing size of the attraction bassin of the $TT$ regime and the decreasing lifetime of the metastable $RT$ regime. For tensions $\alpha \geq 1.2$, $\mathbb{P}_{RT}$ vanishes abruptly: all initial conditions end up in the $TT$ regime at long time. Altogether the active \textit{Gerris} establishes the proof of concept for the experimental control of a switch between two CA regimes using tension.

The \textit{Gerris} structure, which results from several experimental compromises, is however rather artificial. More specifically the fact that it has only three degrees of freedom raises the question of the possible generalization of our results to larger structures. Furthermore the fact that the two branches of eigenfrequencies, corresponding to the TT and RT modes only meet at infinite tension is likely to be specific. 

We now show theoretically that the tension controlled switch is generically expected even in the harmonic approximation. Consider an arbitrary lattice undergoing homogeneous dilation of factor $\alpha \in \left[1,+\infty\right[$, the dynamical matrix of which reads \citep{Supplemental_information}
\begin{equation} \label{eq:homogeneous_dilation}
\mathbb{M}\left(\alpha\right) = \frac{1}{\alpha}\mathbb{M}_0+\left(1-\frac{1}{\alpha}\right) \mathbb{M}_1.
\end{equation}
$\mathbb{M}_0$ is the dynamical matrix of the structure at zero tension, and $\mathbb{M}_1$ 
reads:
\begin{equation}
\mathbb{M}_1 = 
\begin{pmatrix}
\mathbb{M}^{xx}_1 & 0 \\
0 & \mathbb{M}^{yy}_1 \\
\end{pmatrix}.
\end{equation}
where $\mathbb{M}_{1}^{xx} = \mathbb{M}_{1}^{yy}$ is the Laplacian matrix of the structure network $\mathbb{M}_{1,ii}^{\alpha\alpha} = Z\left(i\right)$, $\mathbb{M}_{1,ij}^{\alpha\alpha} = - 1$ if $i$ and $j$ are neighbors and zero otherwise. Since $\mathbb{M}_1$ decouples the $x$ and $y$ directions, its eigenvectors $\varphi_{n}$ come in degenerated pairs with identical form, respectively polarized along $x$ and $y$. In particular, as a result of a discrete nodal domain theorem~\citep{Supplemental_information,duval1999perron,davies2000discrete,gladwell2002courant}, the lowest energy modes of $\mathbb{M}_1$ have the geometry of translational modes. 
Increasing the tension, the spectral properties of $\mathbb{M}_1$ progressively dictate that of the elastic structure, thereby favoring the emergence of two degenerated low energy modes, with geometries akin to translation. These are the perfect conditions for selecting the SLO regime at large tension~\citep{baconnier2022selective}. This is why any elastic structure, which, in the absence of tension, exhibits some form of CA, different from the condensation on modes akin to translation, will eventually switch to the SLO regime, when tension is increased. This argument is strictly valid in the case of a homogeneous dilation, but one expect it to persist as a design principle for CA switch in elastic structures which do not dilate homogeneously, as long as tension is evenly distributed. In the case of the  \textit{Gerris}, the specificity of the spectrum prevents the application of the recipe at finite tension; we however saw that the nonlinearities can enforce it at tensions, that can be reached experimentally. 

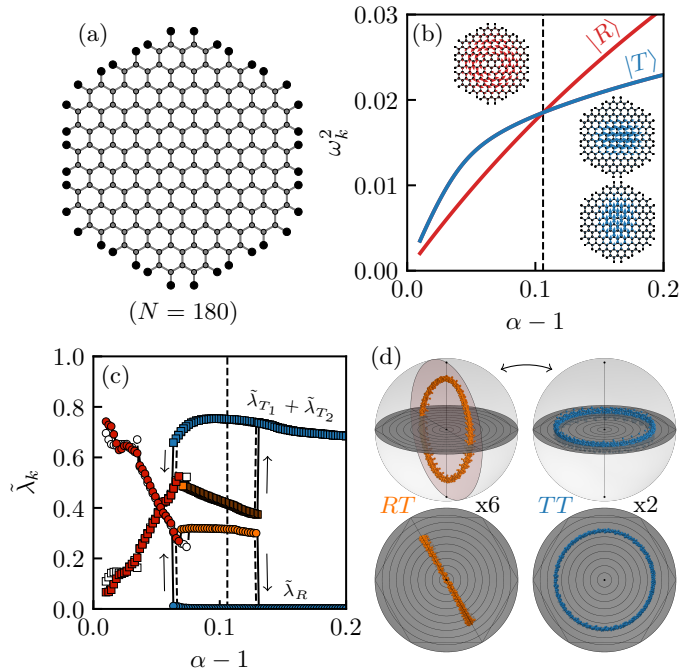
\begin{figure}[t!]
\centering
\hspace*{-0.4cm}
\begin{tikzpicture}

\node[rotate=90] at (0.0,-0.1)
{\includegraphics[height=4.8cm]{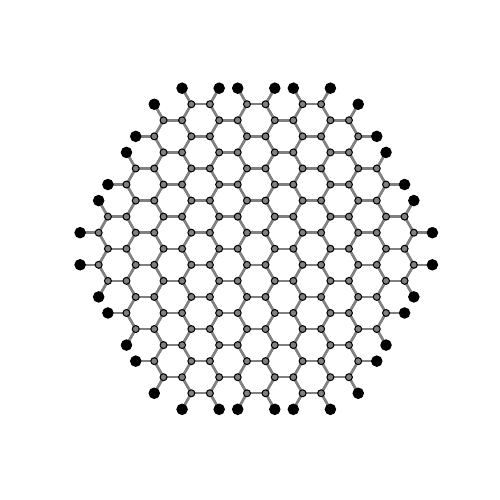}};

\node[] at (0.0,-2.1) {\small ($N = 180$)};

\node[] at (4.2,0.0)
{\includegraphics[height=4.8cm]{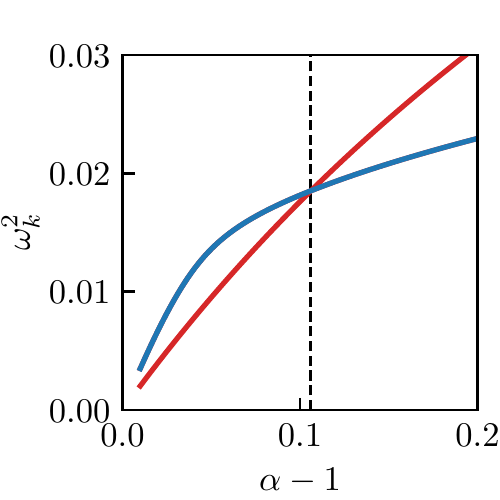}};

\node[rotate=40] at (5.6,1.6) {\small $\color{color2} | R \rangle$};
\node[rotate=15] at (6.1,1.22) {\small $\color{color1} | T \rangle$};

\node[rotate=-90] at (5.77,-0.87)
{\includegraphics[height=1.5cm]{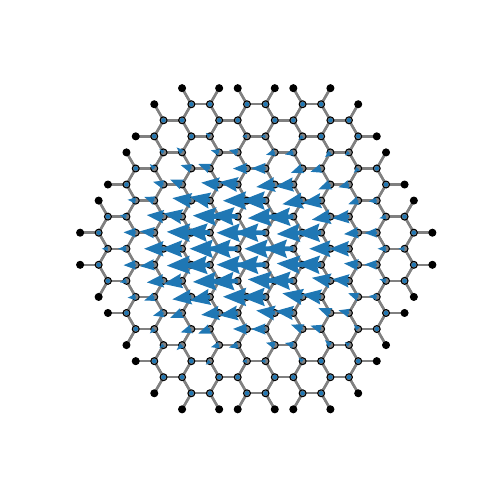}};
\node[rotate=90] at (5.77,0.22)
{\includegraphics[height=1.5cm]{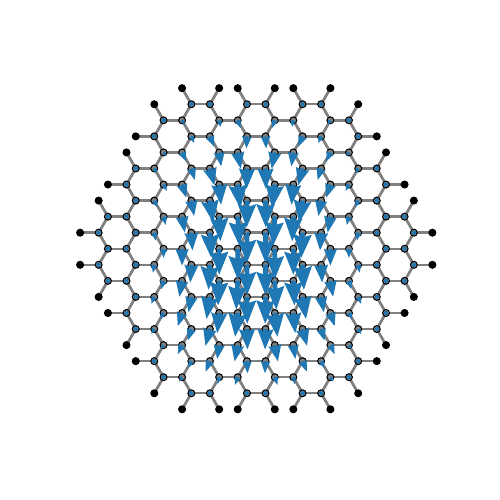}};

\node[rotate=90] at (4.1,1.2)
{\includegraphics[height=1.5cm]{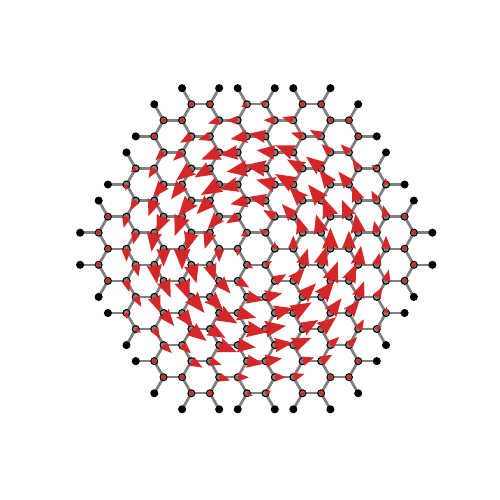}};

\node[] at (0.1,-4.5)
{\includegraphics[height=4.8cm]{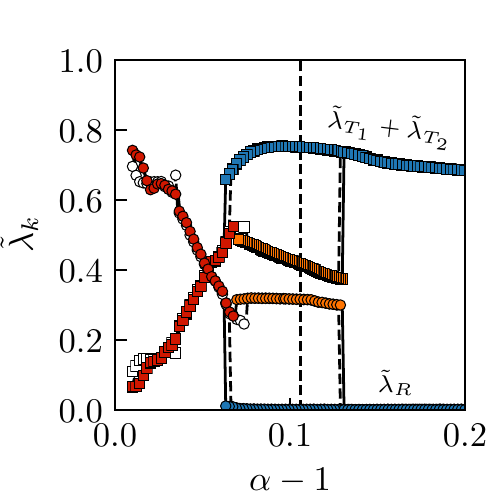}};

\draw[->] (1.11,-5.3) -- (1.12,-5.8);
\draw[->] (-0.25,-5.8) -- (-0.26,-5.3);

\draw[->] (1.11,-4.5) -- (1.12,-4.0);
\draw[->] (-0.24,-3.92) -- (-0.25,-4.22);

\node[] at (3.5,-3.65)
{\includegraphics[height=2.0cm]{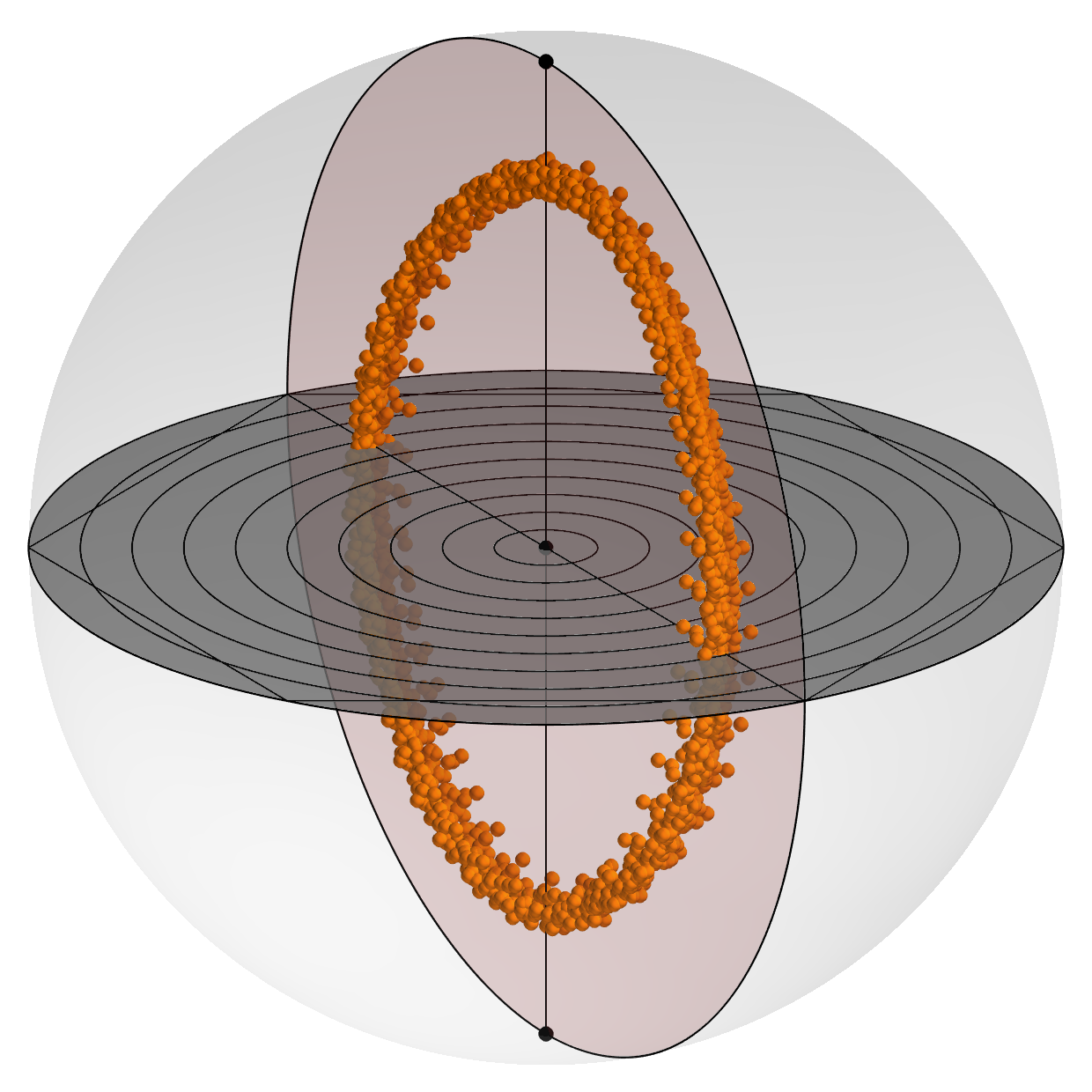}};
\node[] at (5.6,-3.65)
{\includegraphics[height=2.0cm]{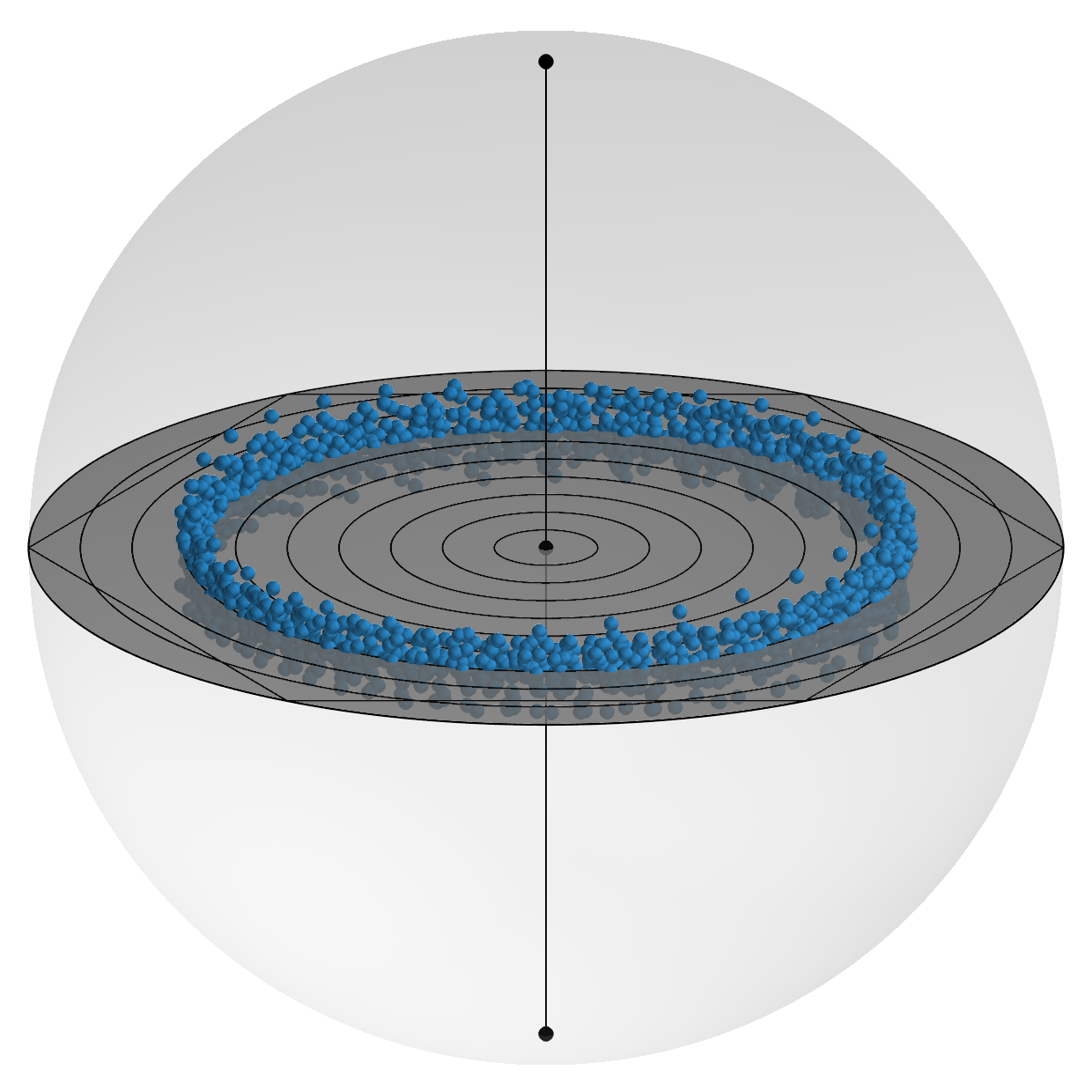}};

\node[rotate=90] at (3.5,-5.65)
{\includegraphics[height=2.0cm]{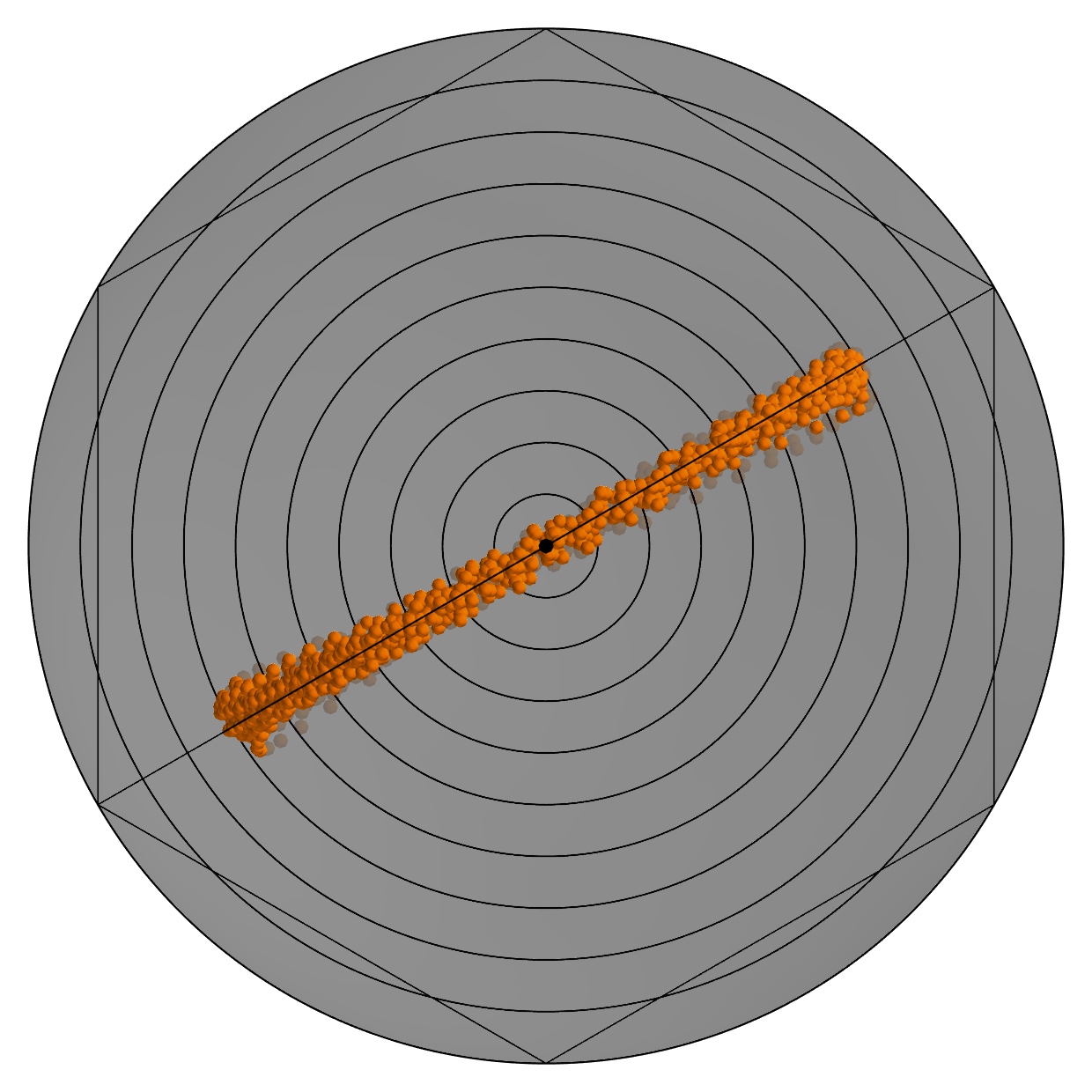}};
\node[rotate=90] at (5.6,-5.65)
{\includegraphics[height=2.0cm]{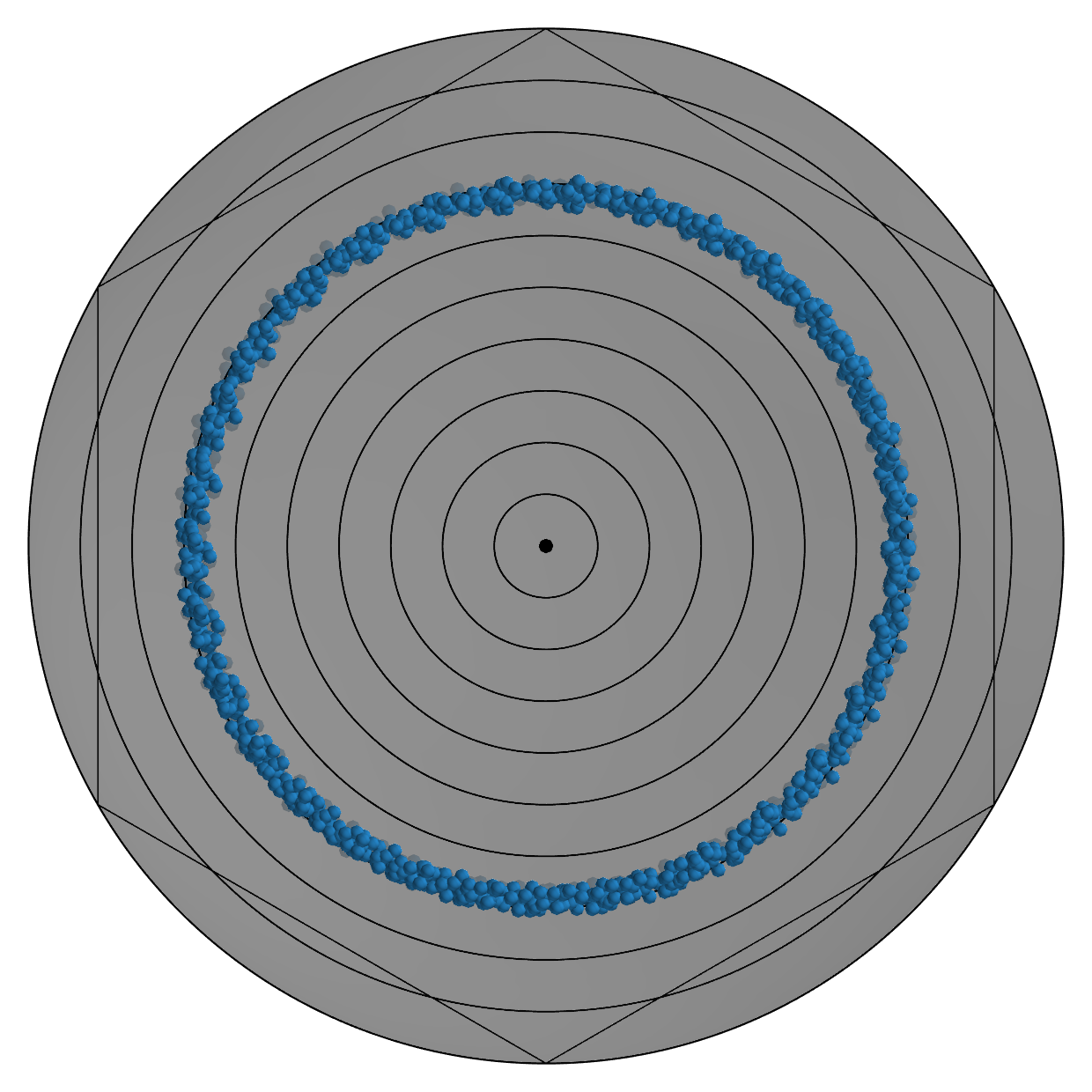}};

\draw[<->] (4.22,-2.85) arc (110:70:1.0);

\node[] at (2.85,-4.65) {\small \color{color_RT1} $RT$};
\node[] at (4.95,-4.65) {\small \color{color1} $TT$};

\node[] at (4.05,-4.65) {\small x$6$};
\node[] at (6.15,-4.65) {\small x$2$};

\node[] at (-1.2,1.6) {\small (a)};
\node[] at (3.25,1.6) {\small (b)};
\node[] at (-0.9,-2.95) {\small (c)};
\node[] at (2.7,-2.7) {\small (d)};

\end{tikzpicture}
\vspace*{-0.7cm}
\caption{\small{\textbf{Active honeycomb's dynamics as a function of lattice tension.} (a) Elastic architecture cartoon ($N = 180$). (b) Normal modes spectrum as a function of lattice tension $\alpha - 1$. The red (resp. blue) solid line corresponds to the rotation mode (resp. degenerated translational modes); see~\cite{Supplemental_information} for the full spectrum. The black dashed vertical line highlights the crossing of energies. (c) Normalized condensation fractions on the rotation mode $\tilde{\lambda}_R$ (circles) and on the translational modes $\tilde{\lambda}_{T_{1}} + \tilde{\lambda}_{T_{2}}$ (squares) as a function of lattice tension $\alpha - 1$. Colored markers and solid lines (resp. white markers and dashed lines) stand for simulations performed within the harmonic approximation (resp. including geometrical non-linearities). The dark red (resp. orange) branch represents the aperiodic (resp. periodic $RT$) GAR regime, while the blue branches represent the $TT$ regime. (d) Projection of the polarity field dynamics in the steady states for the $RT$ (left) and $TT$ (right) regime (vertical axis: $\langle R | \boldsymbol{\hat{n}} \rangle$, equatorial plane: $\langle T_{x/y} | \boldsymbol{\hat{n}} \rangle$).}}
\label{fig:material_scale}
\end{figure}

We confirm this design principle by considering a large honeycomb lattice, composed of $N=180$ nodes, pinned at its hexagonal edges (Fig. \ref{fig:material_scale}-a).  Under small tension this lattice has a rotation mode $| R \rangle$ that lies at the bottom of its vibrational spectrum (Fig.~\ref{fig:material_scale}-b).  As tension increases, the energies of both the degenerated translational modes and the rotation mode increase, but at different pace, and eventually cross each other for $\alpha = \alpha^* \simeq 1.1$, as expected from Eq.~(\ref{eq:homogeneous_dilation}). When simulating the dynamics of the active honeycomb, \emph{within the harmonic approximation}, with $\pi = 0.055$, we do confirm the presence of a tension controlled switch 
between two linearly stable actuation regimes, SLO and GAR (Fig.~\ref{fig:material_scale}-c). Note that the condensation of the dynamics taking place on modes that are not fully delocalized, the condensation fraction are here normalized by the participation ratio of the modes: $\tilde{\lambda}_{k} = \lambda_{k}/Q_k$, with $Q_k = \left( \sum_i | \boldsymbol{\varphi}_k^i | \right)^{2} / N$~\cite{bell1970localization, Supplemental_information}.
The SLO is a $TT$ regime, very similar to the one discussed above (Fig.~\ref{fig:material_scale}-d and SI Movie 5), except for additional fluctuations taking place outside of the equatorial plane. Indeed, the translational modes being not fully delocalized, there is room for a spatial coexistence of a collectively actuated region at the center of the system with a frozen/disordered one close to the boundary. 
The GAR regimes, with strictly positive $\tilde{\lambda}_{R}$, exhibit richer dynamics than in the case of the  \textit{Gerris}~:  for small enough tension, the GAR regimes are aperiodic, because of the many low energy modes, which couple to the rotational and translational modes~\cite{Supplemental_information}. At large enough tension, one recovers the $RT_{1}$ regime, condensed on a $RT$ plane in mode space (Fig. \ref{fig:material_scale}-d and SI Movie 6), modulo some fluctuations of the same origin than in the $TT$ regime.
Annealing from small to large tension, the $RT$ regime switches to the $TT$ one for a tension $\alpha>\alpha^*$ (Fig. \ref{fig:material_scale}-c). Additionally, performing the backward annealing, the $TT$ branch becomes unstable for a tension $\alpha<\alpha^*$. In the absence of geometrical non-linearities, the observed hysteretic switch must be attributed to the non-trivial selectivity of CA. As shown in~\citep{baconnier2022selective}, the selection of the modes ruling the CA not only depends on the energy level, but also on their geometries. More precisely, CA preferably takes places on a pair of modes that are maximally extended and locally orthogonal. 
As demonstrated by the open symbols and dashed lines in Fig. \ref{fig:material_scale}-c, the presence of geometrical non linearities do not alter the above picture.
%In the present case, the geometrical non linearities do not alter the above picture (Fig. \ref{fig:material_scale}-c).

Altogether, having unveiled a new CA regime arising in active solids with a low energy rotation mode (GAR), we demonstrate that mechanical tension is a robust control parameter to switch to a regime dominated by a pair of degenerated translational modes (SLO). On the meta-material science side, our work opens the path towards the study of structures with more low energy modes, and the possible emergence and competition of several actuation branches. In the realm of bio-physics, it suggest the possibility for such switching behavior, for instance in biological tissues, where contractility can generate internal tension.

\begin{acknowledgments}
 We thank Yoav Lahini for useful discussions. We acknowledge financial support from Ecole Doctorale ED564 “Physique en Ile de France” for P.B.’s Ph.D. grant. D.S. was supported by a Chateaubriand fellowship.
\end{acknowledgments}

\bibliographystyle{apsrev4-1} % Tell bibtex which bibliography style to use

\end{document}